\documentclass[a4paper,10pt,english]{article}

\pdfoutput=1

\usepackage{ifpdf,epsfig,array,amsmath,amssymb,psfrag,graphicx,mathrsfs}

\usepackage[colorlinks=true,filecolor=blue,linkcolor=blue,citecolor=blue,urlcolor=blue]{hyperref}
\usepackage[usenames,dvipsnames]{color}

\newcommand{\beq}{\begin{equation}}
\newcommand{\eeq}{\end{equation}}
\newcommand{\bea}{\begin{eqnarray}}
\newcommand{\eea}{\end{eqnarray}}
\newcommand{\qqquad}{\quad\qquad}

\newcommand{\intd}{\mathrm{d}}

\newcommand{\mc}[1]{\mathcal{#1}}

\newcommand{\ii}{\mathrm{i}}
\newcommand{\ee}{\mathrm{e}}

\newcommand{\Real}{\mathrm{Re}\, }
\newcommand{\Img}{\mathrm{Im}\, }

\newcommand{\scri}{{\mathscr{I}}}
\newcommand{\scrip}{\scri^{+}}

\makeatletter \@addtoreset{equation}{section} \makeatother 
\renewcommand{\theequation}{\arabic{section}.\arabic{equation}}

\begin{document}

\title{Numerical investigation of the late-time tails of the solutions of the Fackerell--Ipser equation}

\author{Istv\'an R\'acz\thanks{racz.istvan@wigner.hun-ren.hu} \ and G\'abor Zsolt T\'oth\thanks{toth.gabor.zsolt@wigner.hun-ren.hu}\\[2mm]
\small \it HUN-REN Wigner RCP, H-1121 Budapest, Konkoly-Thege Mikl\'os \'ut 29-33, Hungary}

\date{}

\maketitle

\begin{abstract}
The late-time behaviour of the solutions of the Fackerell--Ipser equation
(which is a wave equation for the spin-zero component of the electromagnetic field strength tensor)
on the closure of the domain of outer communication of sub-extremal Kerr spacetime is studied numerically.
Within the Kerr family, the case of Schwarzschild background is also considered.
Horizon-penetrating compactified hyperboloidal coordinates are used,
which allow the behaviour of the solutions to be observed at the event horizon
and at future null infinity as well.
For the initial data, pure multipole configurations that have compact support
and are either stationary or non-stationary are taken.
It is found that with such initial data
the solutions of the Fackerell--Ipser equation converge at late times
either to a known static solution (up to a constant factor) or to zero.
As the limit is approached, the solutions exhibit a quasinormal ringdown and finally a power-law decay.
The exponents characterizing the power-law decay of the spherical harmonic components of the field variable
are extracted from the numerical data for various values of the parameters of the initial data,
and based on the results a proposal for a Price's law relevant to the Fackerell--Ipser equation is made.
Certain conserved energy and angular momentum currents are used to verify
the numerical implementation of the underlying mathematical model.
In the construction of these currents a discrete symmetry of the Fackerell--Ipser equation,
which is the product of an equatorial reflection and a complex conjugation,
is also taken into account.
\end{abstract}

\vspace{1cm}

\newpage

\section{Introduction}
\label{sec.intr}

The perturbations of Kerr black holes have been studied extensively over the past few decades
because of their importance for general relativity and astrophysics.
Although much is now known about them, their study is not yet complete.

The most commonly used equation in the study of the perturbations of Kerr black holes is the Teukolsky master equation (TME).
This is a wave equation for the extreme spin-weight Newman--Penrose components
(with respect to the Kinnersley tetrad) of the electromagnetic field strength
or the linearized Weyl tensor, depending on the value of the spin parameter.
(Note that certain components of the spin-$1/2$ and spin-$3/2$ fields
are also solutions of the TME with the corresponding value of the spin parameter \cite{Teukolsky1,Teukolsky2,CasOrt,Ort}).
Nevertheless, the other components of the electromagnetic field strength and the linearized Weyl tensor
also satisfy analogous wave equations \cite{FI,AndAks,Aksteiner}.
Notably, the spin-zero component of the electromagnetic field strength satisfies the Fackerell--Ipser (F--I) equation \cite{FI}.
Like the TME, this equation is distinguished in that it is decoupled from
the wave equations relevant to other components of the perturbing fields.

As discussed in \cite{FI}, once the solution to the F--I equation is known,
the other components of the perturbing electromagnetic field can be determined algebraically or by quadrature.
The F--I equation also played an important role in \cite{AndBlu},
where energy and Morawetz estimates were derived for both the full Maxwell equations and the F--I equation in the exterior of very slowly rotating Kerr black holes.
The estimates were used in \cite{AndBlu} to prove a uniform bound on a positive definite energy
and the convergence of the Maxwell field to a static Coulomb field.
(For results relevant to the Schwarzschild background, see also \cite{Blue,Ghanem,SteTat,AndBacBlu,Pasqualotto}).

In the present work, our main goal is to numerically study the late-time behaviour of the solutions of the Fackerell--Ipser equation
on the exterior of sub-extremal Kerr spacetime,
using a framework that incorporates the techniques of conformal compactification and the hyperboloidal initial value problem.
The time slices in the latter are chosen to be horizon penetrating, allowing us to study the behaviour of the solutions at the event horizon,
in addition to future null infinity and the locations at finite distance from the event horizon.
We have also used this framework in \cite{RT,CRT},
where we studied the late-time tails of the solutions of the TME and the scalar wave equation.

After a sufficiently long evolution,
the solutions of the TME and the scalar wave equation generated from pure multipole initial data
that decrease sufficiently rapidly as infinity is approached
exhibit a power-law decay in time (i.e., they depend on time asymptotically as $\sim t^{-n}$,
where $n$ is some positive integer).
The exponent characterizing the decay depends on some parameters of the initial data
and can take different values at the event horizon, at finite distance from the event horizon and at future null infinity.
The solutions of the F--I equation are also expected to show such behaviour,
although some solutions may converge to a static configuration instead of zero, as indicated by the results of \cite{AndBlu}.
In the latter case, the power-law behaviour can be expected to be exhibited by the part of the solution obtained by subtracting the static part.

Our aim concerning the late-time behaviour of the solutions of the Fackerell--Ipser equation
is to study the power-law decay of the multipole components of the solutions,
and to determine the rate of their decay
at the event horizon, at finite distance from the event horizon, and at future null infinity
as a function of the parameters of the initial data and the indices of the multipole components.

We also aim to verify whether the above-mentioned phenomenon of convergence of some solutions
to a static configuration occurs.
We are primarily interested in the case of the Kerr background with moderate angular velocity.
However, because of its particular interest, investigations of the case of the pure Schwarzschild background will also be included.

As in our previous work \cite{RT,CRT}, we use certain conserved energy and angular momentum currents
to verify the numerical implementation of the underlying mathematical model.
We construct these currents in an analogous way as in the case of the TME \cite{Toth},
but also taking into account a discrete symmetry of the F--I equation,
which is the product of an equatorial reflection and a complex conjugation.

The stability of Kerr black holes under linear perturbations,
in particular the late-time tails of the solutions of the TME and the scalar wave equation,
has been studied very extensively by other authors in \cite{Whiting}-\cite{Millet},
both analytically and numerically, also in both the frequency and time domains.
(For non-rotating black holes, see, e.g., \cite{Price1,Price2}, \cite{RegWhe}-\cite{AngAreGaj2}).
Specifically for the F--I equation, however, not many results are available;
to our knowledge, the present study is the first numerical one,
while on the analytical side only \cite{AndBlu} is known to us.

The pioneering results on the power-law decay of linear fields on black hole spacetimes were due to Price \cite{Price1,Price2}.
He showed that a generic initially compactly supported radiative $l$ multipole component of a massless perturbative field of integer spin
in the Schwarzschild background dies out at late times as $t^{-(2l+3)}$.
(The radiative multipoles are those with $l\ge s$, where $s$ is the spin of the perturbative field).
On the Kerr background, analytical results extending those of Price
were first obtained by Barack and Ori \cite{BarOri,Barack,BarOri2}, and Hod \cite{Hod1,Hod2,Hod3}.
Casals et al.\ \cite{CasKavOtt,CasOtt} also determined higher order terms.
The first numerical investigations on the Kerr background were carried out by Krivan et al.\ \cite{KriLagPap,KriLagPapAnd,Krivan}.

Our paper is organized as follows.
In section \ref{sec.fieq}, the F--I equation is introduced.
In section \ref{sec.method}, the methods we use to solve the F--I equation are described.
In particular, the coordinate system is specified, the multipole expansion of the field variable is introduced,
and the F--I equation is rewritten in a form suitable for the numerical computation of the time evolution of the expansion coefficients.
Some of the details are moved to appendices \ref{app.coeff} and \ref{app.lightspeed}.
In section \ref{sec.sym},
the discrete symmetry of the F--I equation mentioned above is described
and the energy and angular momentum type currents are constructed.
In section \ref{sec.result}, the initial data are specified and the results of our numerical calculations
on the late-time power-law decay of the solutions of the F--I equation are presented.
An account of the test of the conservation of the currents $\mc{E}_P^\mu$ and $\mc{J}_P^\mu$,
introduced in section \ref{sec.sym}, is also given. Some technical details for the latter test are collected in appendix \ref{app.cl}.
Section \ref{sec.concl} contains our concluding remarks.

\section{The Fackerell--Ipser equation}
\label{sec.fieq}

We recall that 
the line element of the Kerr metric in Boyer--Lindquist coordinates $(t,r,\theta,\phi)$ reads 
\beq
\label{eq.metric}
ds^2=\left(1-\frac{2Mr}{\Sigma}\right)dt^2+\frac{4 a r M\sin^2\theta}{\Sigma}dt d\phi
-\frac{\Sigma}{\Delta}dr^2 - \Sigma d\theta^2 -\frac{\Gamma }{\Sigma}\sin^2\theta d\phi^2,
\eeq 
where
$\Sigma=r^2+a^2\cos^2\theta$, $\Delta=r^2-2Mr+a^2$, $\Gamma=(r^2+a^2)^2-a^2 \Delta\sin^2\theta$.
The parameters $M$ and $a$ are the mass and
the angular momentum per unit mass of the black hole contained by the Kerr spacetime.
(\ref{eq.metric}) implies the signature $(+,-,-,-)$ for the metric.
The Kinnersley null tetrad is given by
\beq
l^\mu=\frac{1}{\Delta}(r^2+a^2,\Delta,0,a),\qquad
n^\mu = \frac{1}{2\Sigma}(r^2+a^2,-\Delta,0,a),
\eeq
\beq
m^\mu = \frac{1}{\sqrt{2}(r+\ii a\cos\theta)}\left(\ii a \sin\theta,0,1,\frac{\ii}{\sin\theta}\right).
\eeq

The Fackerell--Ipser equation \cite{FI} is a wave equation
for the spin-weight zero Maxwell scalar 
\beq
\phi_1  =  \frac{1}{2} F_{\mu\nu}(l^\mu n^\nu - m^\mu \bar{m}^\nu),
\eeq
where $F_{\mu\nu}$ is the electromagnetic field, in Kerr spacetime.
It takes the form
\beq
\label{eq.fi}
\nabla^\mu \nabla_\mu \Omega +2\Psi_2 \Omega = 0,
\eeq
where
\beq
\Omega  = \phi_1/\rho\, ,\quad
\Psi_2  = M\rho^3,\quad
\rho= -(r-\ii a\cos\theta)^{-1}\, .
\eeq
$\Psi_2$ is the spin-weight zero Weyl scalar of the Kerr spacetime
($\Psi_2 =  -C_{\mu\nu\lambda\rho}l^\mu m^\nu \bar{m}^\lambda n^\rho$, where
$C_{\mu\nu\lambda\rho}$ denotes the Weyl tensor of the Kerr spacetime).
Note that $\Omega$ is a complex variable, and the potential $\Psi_2$ is also complex.

The F--I equation has two well-known exact solutions, a static one,
\beq
\label{eq.omst}
\Omega_{\mathrm{st}} \, =\, \frac{1}{r-\ii a \cos\theta}\, ,
\eeq
and a linearly growing one,
\beq
\Omega_{\mathrm{lin}} \, =\, t\, \Omega_{\mathrm{st}}\, .
\eeq
According to \cite{FI}, the F--I equation is not separable, in contrast with the Teukolsky master equation.

Although in this paper our attention will be restricted to the case where the gravitational background is the Kerr spacetime,
we mention that the F--I equation has been generalized to vacuum type D spacetimes \cite{CroFac}-\cite{Araneda}.

\section{The applied methods}
\label{sec.method}

In this section the methods used in numerically solving the F--I equation are discussed, 
along with additional properties of the F--I equation. Some of the details are moved to Appendix
\ref{app.coeff}.

\subsection{Horizon-penetrating compactified hyperboloidal coordinates}
\label{sec.coords}

We use the same coordinates, with a very minor modification, as in \cite{RT, CRT}.
Concerning the literature on the ideas and techniques underlying these coordinates and the rewriting of the F--I equation
in Section \ref{sec.numm}, we refer the reader to \cite{Penrose1}-\cite{Zenginoglu}, \cite{ZenTig, ZenNunHus, Jasiulek, HBB, MacAns,
MacJarAns, Macedo} and references therein.

\begin{figure}[h]
%
\begin{center}
\includegraphics[scale=0.5]{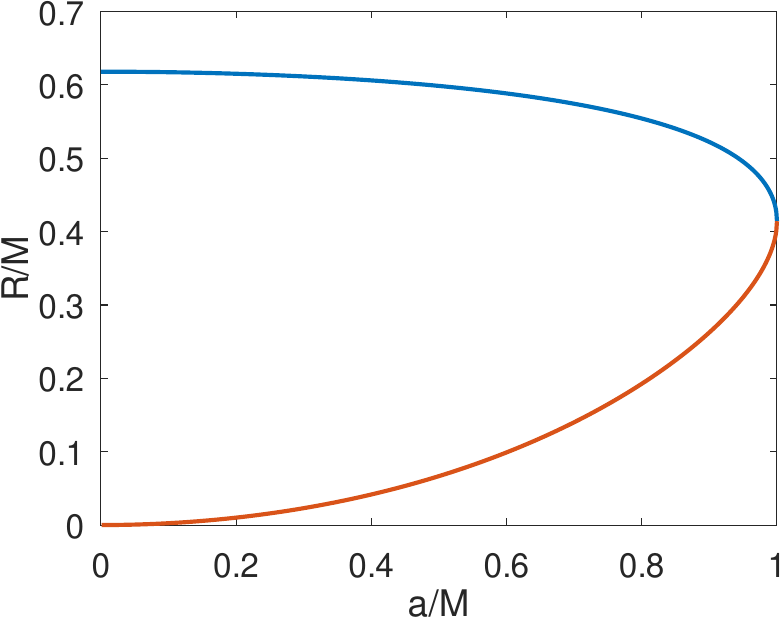} 
\end{center}
\vspace{-0.5cm}
\caption{\label{fig.horizR} Location of the event horizon (blue curve) and the inner horizon (red curve) as a function of $a/M$}
\end{figure}

The horizon-penetrating compactified hyperboloidal coordinates are obtained from the Boyer--Lindquist coordinates by two transformations.
The first one is the transformation to ingoing Kerr coordinates $(\tau,r,\theta,\varphi)$: 
\beq
\tau  =  t-r+\int\intd r\, \frac{r^2+a^2}{\Delta}\, , \qqquad 
\varphi  =  \phi+ \int\intd r\, \frac{a}{\Delta}\, . 
\eeq
The location of the event horizon and the inner horizon is at
$r=r_{\pm}=M\pm\sqrt{M^2-a^2}$ in these coordinates.
In the second transformation we replace $\tau$ and $r$ by the new time coordinate $T$ and by the compactified radial coordinate $R$,
defined via the implicit relations 
\begin{eqnarray}
\label{eq.tr3}
\tau & = & T+M\frac{M^2+R^2}{M^2-R^2}-4M\log(|1-(R^2/M^2)|)\\
\label{eq.tr4}
r & = & \frac{2R}{1-(R^2/M^2)}\,.
\end{eqnarray}
The value of $R$ at the event horizon will be denoted by $R_+$.
The value of $R/M$ at the event horizon and at the inner horizon as a function of $a/M$ is shown in Figure \ref{fig.horizR}.
Future null infinity ($\scrip$) is located at $R=M$ in the coordinates $(T,R,\theta,\varphi)$.
The $T=const$ surfaces tend to future null infinity as $r\to\infty$, i.e., as $R\to M$.
They are spacelike in the domain of outer communication and remain spacelike
in the interior of the black hole between the inner horizon and the event horizon.

\begin{figure}[h]
%
\begin{center}
\hspace*{1cm}$a/M=0.5$\hspace{6cm}$a/M=0.5$\\[2mm]
\includegraphics[height=4.8cm]{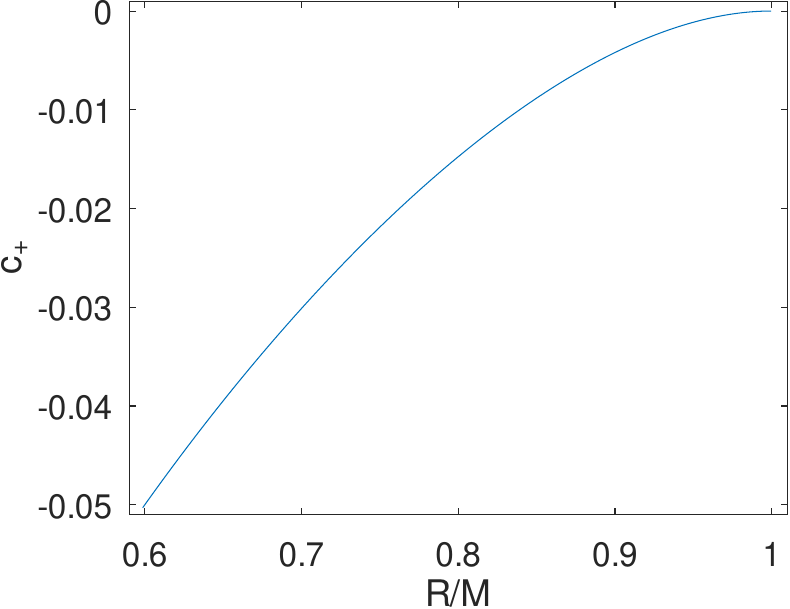} 
\hspace{1cm}
\includegraphics[height=4.8cm]{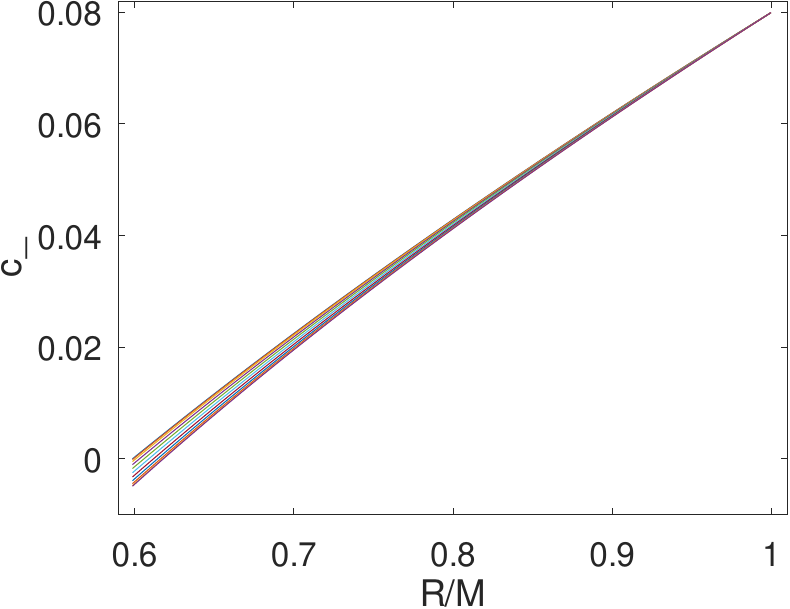}\\[1mm]   
\hspace{0.9cm}(a)\hspace{7cm}(b)
\end{center}
\vspace{-0.5cm}
\caption{\label{fig.cpm05}
(a) The ingoing speed of light $c_+$ as a function of $R/M$ in the range $[R_+/M,1]$ at $a/M=0.5$.
(b) The outgoing speed of light $c_-$ as a function of $R/M$ in the range $[R_+/M,1]$ at $a/M=0.5$
and $\theta=(\pi/2)(n/10)$, $n=0,1,\dots, 10$}
\end{figure}

Although some of the coordinate components of the metric tensor are singular at $R=M$, the conformally rescaled
metric $r^{-2}g_{\mu\nu}$ is regular.
It is straightforward to see that the coordinate basis fields $(\partial_T)^\mu$ and $(\partial_\varphi)^\mu$ are Killing vector fields.
In comparison with \cite{RT,CRT}, we have included some factors of $M$ in the transformation (\ref{eq.tr3}), (\ref{eq.tr4}),
in order to respect the physical dimension of the various terms. Without these factors the transformation contains implicitly a 
second mass scale beside $M$.
In the numerical calculations we set $M=1$,
as we did in \cite{RT,CRT}, and for this value the transformation applied in \cite{RT,CRT}
coincides with the one above.

A major advantage of the coordinates ($T,R,\theta,\varphi$) is that
they allow one to place the inner boundary of the computational domain inside the black hole region
and the outer boundary to future null infinity.
In our numerical calculations, the inner boundary is at a constant value of $R$ that is somewhat smaller than $R_+$
and the outer boundary is at $R=1$.
The $R=const$ surfaces are spacelike between the event horizon and the inner horizon,
thus the inner boundary is a spacelike surface.
The light cones with vertices inside the computational domain and at the boundaries intersect
the Cauchy surface in compact subsets,
so that the field values at these points are uniquely determined by the initial data.
At the same time, none of the future light cones with vertices at the boundary points
intersect the interior of the computational domain; hence, the boundary surfaces allow waves to leave this domain.
Therefore it is not necessary to impose any boundary conditions;
it is sufficient to simply require that the field equation holds at the edges of the computational domain,
which allows a significant reduction of spurious boundary effects
(see e.g.\ \cite{DafRod-artb} for a discussion on the effects of boundary conditions).

For an illustration of the direction and speed of wave propagation
in the coordinate system $(T,R,\theta,\varphi)$,
see the plots of the ingoing and outgoing radial lights speeds in the domain of outer communication
in Figure \ref{fig.cpm05}.
These light speeds are discussed in detail in Appendix \ref{app.lightspeed},
where additional figures can also be found.

\subsection{The numerical method}
\label{sec.numm}

For the purpose of numerical solution, we rewrite the F--I equation in the form 
\bea
\label{eq.de}
\partial_{T}^2\Psi & = & \frac{1}{a_{TT}}(a_{RR}\,\partial_{R}^2+a_{TR}\,\partial_{TR} 
+a_{T\varphi}\,\partial_{T\varphi}+a_{R\varphi}\,\partial_{R\varphi} \nonumber \\ 
&&\hspace{1cm}
+\,a_{T}\,\partial_{T}+a_{R}\,\partial_{R}+a_{\varphi}\,\partial_{\varphi}
+a_0 +a_{\Delta}\Delta_{S^2})\,\Psi
-\frac{2Q}{ra_{TT}}\Psi_2\Psi\, , 
\eea
where
\bea
\label{eq.resc}
\Psi & = & r\,\Omega\, ,\\
a_{TT} & = & a_{TT}^{(0)}+a_{TT}^{(2)}\,Y_2^0\, , \\
Q & = & Q^{(0)}+Q^{(2)}\,Y_2^0\, ,
\eea
$Y_2^0=\frac{1}{4}\sqrt{\frac{5}{\pi}}(3\cos^2\theta-1)$ is the spherical harmonic function 
with $l=2$, $m=0$, and $\Delta_{S^2}$ stands for the spherical Laplace operator
$\frac{1}{\sin\theta}\,\frac{\partial}{\partial \theta}\left(\sin\theta\,\frac{\partial}{\partial\theta}\right)
+ \frac{1}{\sin^2\theta}\, \frac{\partial^2}{\partial \varphi^2}$.

The coefficients $a_{RR}$, $a_{TR}$, $a_{T\varphi}$, $a_{R\varphi}$, $a_T$, $a_R$, $a_\varphi$,
$a_0$, $a_{\Delta}$, $a_{TT}^{(0)}$, $a_{TT}^{(2)}$, $Q^{(0)}$ and $Q^{(2)}$ depend on $R$ but not on $T$, $\theta$, $\varphi$.
Their explicit form is given in Appendix \ref{app.coeff}. We note that $g^{TT}=r\,a_{TT}/Q$.
$\Psi_2$ can be expressed as 
\beq
\label{eq.psi2}
\Psi_2=-\frac{M(M^2-R^2)^3}{8\bigl(RM^2-\ii a(M^2-R^2)\sqrt{\frac{\pi}{3}}Y_1^0\bigr)^3}\, ,
\eeq
where $Y_1^0=\frac{1}{2}\sqrt{\frac{3}{\pi}}\cos\theta$ is the spherical harmonic function with indices $l=1$, $m=0$.
$a_{RR}$, $a_{TR}$, $a_{T\varphi}$, $a_{R\varphi}$, $a_T$, $a_R$, $a_\varphi$,
$a_0$, $a_{\Delta}$, $a_{TT}^{(0)}$, $a_{TT}^{(2)}$, $1/a_{TT}$ and $\Psi_2$ are regular in the exterior region,
including the event horizon and the boundary at $R=M$.
Although $Q$ contains a singular factor $1/(M-R)^3$, this is cancelled out by a factor $(M-R)^3$ in $\Psi_2$, therefore
the last term in (\ref{eq.de}) is also regular.
The rescaling (\ref{eq.resc}) is needed for the regularity of the wave equation at $R=M$.

Our approach to the numerical solution of (\ref{eq.de}) is essentially the same as in \cite{RT}.
The angular dependence of the field is handled by means of spectral decomposition,
i.e.\ $\Psi$ is expanded in the series of spherical harmonics
\beq
\label{eq.ser}
\Psi(R,T,\theta,\varphi)=\sum_{l=0}^\infty \sum_{m=-l}^l \psi_l^m(R,T)\,Y_l^m(\theta,\varphi)\,,
\eeq
which converts the evolution problem into one for the coefficients $\psi_l^m(R,T)$
(these coefficients, or the terms on the right hand side of (\ref{eq.ser}), are often referred to as \textit{projected modes} in the literature).
In practice, the series (\ref{eq.ser}) has to be truncated to a finite sum,
nevertheless, as it converges quickly, the desired precision can always be achieved by keeping sufficiently many terms.
The evaluation of the right hand side of (\ref{eq.de}) with regard to the expansion into spherical harmonics is done in the same 
way as in \cite{RT}, in completely algebraic manner. 
The additional last term in (\ref{eq.de}) does not require any new techniques. The action of 
$\Delta_{S^2}$ and $\partial_\varphi$ can be evaluated using the properties 
$\Delta_{S^2}Y_l^m=-l(l+1)Y_l^m$ and $\partial_\varphi Y_l^m=\ii mY_l^m$ of the spherical harmonic functions.
The division by $a_{TT}$ and by the denominator of $\Psi_2$ (see (\ref{eq.psi2}))
can be converted into multiplications by applying the identity
\beq
\label{eq.11x}
1/(1+x)= \sum_{k=0}^\infty (-x)^k\, ,
\eeq
with $x=a_{TT}^{(2)}Y_2^0/a_{TT}^{(0)}$ and $x=-\ii a (M^2-R^2)\sqrt{\pi/3}\,Y_1^0/(RM^2)$, respectively.
Multiplications by spherical harmonics can be executed using the relation
\beq
Y_{l_1}^{m_1}Y_{l_2}^{m_2}=\sum_{l_3}G_{l_1l_2l_3}^{m_1m_2m_3}Y_{l_3}^{m_3}\, ,
\eeq
where $m_3=m_1+m_2$, and $G_{l_1l_2l_3}^{m_1m_2m_3}$ are the Gaunt coefficients for which explicit formulae
are available in standard mathematical references.  
These techniques for multiplication and division are also explained in detail in \cite{CLR}.

\begin{figure}[h]
%
\begin{center}
\includegraphics[scale=0.55]{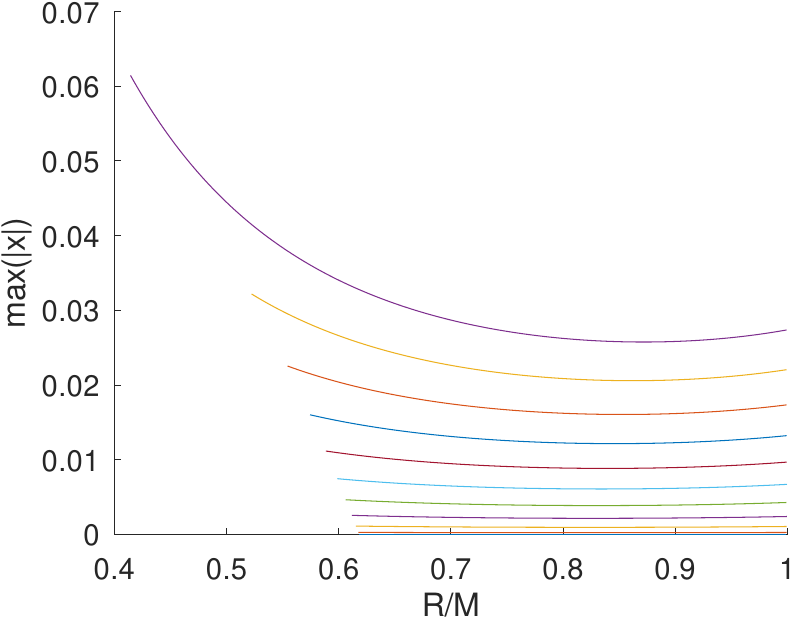} 
\end{center}
\vspace{-0.5cm}
\caption{\label{fig.xatt} $\mathrm{max}(|x|)=a_{TT}^{(2)}\mathrm{max}(|Y_2^0|)/a_{TT}^{(0)}$ as a function of $R/M$ for $a/M=0,0.1,0.2,\dots ,1$
in the exterior region of the Kerr spacetime.
$\mathrm{max}(|Y_2^0|)=\frac{1}{2}\sqrt{\frac{5}{\pi}}$
}
\end{figure}

In the case of $a_{TT}$, $|x| < 0.00775$ for $a/M=0.5$, which is the value $a/M$ takes in almost all of our simulations.
Figure \ref{fig.xatt} shows upper bounds on $|x|$ as a function of $R/M$ at $a/M=0,0.1,0.2,\dots,1$. These bounds increase with $a/M$,
but they are small even at $a/M=1$. 
Hence, the series on the right hand side of (\ref{eq.11x}) converges rapidly,
and only a few terms of it need to be kept to achieve sufficient precision.

In the case of the denominator of $\Psi_2$,
$|x|\le a/r$. In particular, $|x|\le a/(M+\sqrt{M^2-a^2})$ at the event horizon. This quantity approaches $1$ as $a/M\to 1$, 
showing that the number of terms that need to be kept in (\ref{eq.11x}) to achieve sufficient precision
becomes large for nearly extremal black holes.
Nevertheless, for 
$a/M=0.5$, $|x|< 0.268$ in the exterior region.

Since the F--I equation is linear and possesses the axial symmetry of the Kerr background as well,
the time evolution of the coefficients $\psi_l^m(R,T)$ with different azimuthal ($m$) indices decouples,  
allowing one to investigate their evolution separately.
The speed of the numerical calculations can also be increased significantly by  
taking into consideration the decoupling of 
$\psi_l^m(R,T)$ with different values of $m$.

For the numerical solution, (\ref{eq.de}) is rewritten in first order form by introducing 
the additional dependent variables $\Psi_T=\partial_T\Psi$ and $\Psi_R=\partial_R\Psi$.
The method of lines is applied in the $T - R$ plane to compute the time evolution 
of the multipole coefficients of $\Psi$, $\Psi_T$ and $\Psi_R$.
Fourth order finite differencing is used in the radial direction, while for the time evolution
a fourth order Runge--Kutta scheme is applied.
A standard fifth order Kreiss--Oliger dissipation term, as proposed in \cite{Gustetal}, is added
to the right hand side of the equations in order to suppress high frequency instabilities, 
and the constraint $\Psi_R=\partial_R\Psi$ is imposed after each time step.
The time step is chosen to be ten times the radial lattice spacing in accordance with the 
smallness of the coordinate light speed (see Appendix \ref{app.lightspeed} for a discussion of radial light speeds).
The number of grid cells in the radial direction is taken to be $2048$ or $4096$ in most of the simulations,
as this gives sufficiently precise results without making the computation time too long.

\section{Symmetries of the Fackerell--Ipser equation}
\label{sec.sym}

In this section some discrete symmetries of the Fackerell--Ipser equation in the $(T,R,\theta,\varphi)$ coordinates are discussed,
and conserved currents associated with time translation and axial rotation symmetry are constructed. 
We used these conserved currents, combined with a discrete symmetry (denoted by $P$; see (\ref{eq.P}) below),
to test our code for solving the Fackerell--Ipser equation;
for further details concerning this test, see Section \ref{sec.cctest} and Appendix \ref{app.cl}.
$P$ also explains a decoupling of the time evolution of the real and imaginary parts of the coefficients
$\psi_l^0(R,T)$,
and it relates initial data (specified in Section \ref{sec.initc}) that have azimuthal indices of opposite sign.
The latter relation is quite useful as it shows that
it is sufficient to investigate the time evolution of initial data with non-negative azimuthal index.

Other recent works, beside \cite{Toth,CRT}, where partly similar currents
appear in relation to the perturbations of Kerr spacetime
are \cite{Aksteiner}, \cite{ABl}-\cite{GreHolSbeTooZim}.

\subsection{Discrete symmetries}
\label{sec.discrete}

As mentioned in Section \ref{sec.numm}, the $\psi_l^m(R,T)$ coefficients with different $m$ index evolve independently.
The scalar wave equation ($\nabla_\mu \nabla^\mu \Phi=0$) is also
invariant under the equatorial reflection symmetry of the Kerr spacetime,
resulting in the further decoupling of the $\psi_l^m(R,T)$ according to the parity of their $l$ index. 
In the case of the F--I equation, however, the latter decoupling does not take place,
since the additional potential term in the F--I equation violates the equatorial reflection symmetry.
Regarding the structure of (\ref{eq.de}), 
the mixing between different spherical harmonic modes is caused by the $Y_2^0$ term in $a_{TT}$ and $Q$
and by the $Y_1^0$ term in $\Psi_2$.
In particular, the mixing between modes with $l$ indices of different parity is caused by the $Y_1^0$ term in $\Psi_2$.
In the Schwarzschild limit these terms are absent,
and the evolution of the $\psi_l^m(R,T)$ coefficients completely decouple.

Another discrete symmetry of the scalar wave equation is the complex conjugation, which results in the decoupling of the 
real and imaginary parts of $\psi_l^0(R,T)$. This symmetry is also missing in the case of the F--I equation, 
except in the Schwarzschild limit, because $\Psi_2$ is not real if $a\ne 0$.

Nevertheless, although the equatorial reflection and the complex conjugation are not symmetries of the F--I equation, their product
\beq
\label{eq.P}
P: \Psi(T,R,\theta,\varphi)\to \Psi(T,R,\pi-\theta,\varphi)^*
\eeq
is, since $P\Psi_2=\Psi_2$. 
The action of $P$ on the spherical harmonic functions is 
\beq
\label{eq.PY}
Y_l^m(\pi-\theta,\varphi)^* = (-1)^l Y_l^{-m}(\theta,\varphi)\, .
\eeq

As can be seen from (\ref{eq.P}) and (\ref{eq.PY}),
the symmetry of the F--I equation under the action of $P$ implies a decoupling in the axially symmetric ($m=0$) sector:
$\Real \psi_l^0(R,T)$ is coupled to $\Real \psi_{l'}^0(R,T)$   
and
$\Img \psi_l^0(R,T)$ is coupled to $\Img \psi_{l'}^0(R,T)$ 
only if $l-l'$ is even, 
and $\Real \psi_l^0(R,T)$ and $\Img \psi_{l'}^0(R,T)$ are coupled only if $l-l'$ is odd.

To summarize, although the F--I equation with $a\ne 0$ lacks some of the discrete symmetries that the scalar wave equation has,
it is invariant under the action of $P$, which implies the decoupling described in the previous paragraph.
If $a=0$, then the F--I equation has equatorial reflection symmetry and complex conjugation symmetry as well.

We note that the Teukolsky master equation also has a discrete symmetry acting in the same way as $P$
(see Section 3 of \cite{Toth}).

\subsection{Conserved currents associated with time translations and rotations}
\label{sec.cc}

The same considerations that led to the Lagrangian (3.12)
in the case of the Teukolsky master equation in \cite{Toth}
lead in the case of the F--I equation to the result that
\beq
\label{eq.l}
L= \sqrt{-g}\, \bigl(-(\nabla_\mu \Omega_1) (\nabla^\mu \Omega_2) + 2 \Psi_2 \Omega_1\Omega_2\bigr)\, ,
\eeq
where $g$ denotes the determinant of the metric of the background and
$\Omega_1$ and $\Omega_2$ are complex scalar field variables,
is a (complex) Lagrangian density for two independent copies of the F--I equation
(namely, $\nabla^\mu\nabla_\mu \Omega_{1,2}+2\Psi_2\Omega_{1,2}=0$).
Time translations and rotations around the axis of the black hole are Noether symmetries of
$L$, therefore one can obtain conserved currents associated with these symmetries by applying Noether's theorem. 
The (first order) variations of $\Omega$ under time translations and rotations are
$\partial_T \Omega$, $\partial_\varphi\Omega$,
and the corresponding variations of $L$ are $\partial_T L$, $\partial_\varphi L$.
The associated Noether currents are thus
\bea
\mc{E}^\mu [\Omega_1, \Omega_2] & = &   (\nabla^\mu \Omega_1) (\nabla_T \Omega_2) + (\nabla^\mu \Omega_2) (\nabla_T \Omega_1)
+ \delta^\mu_T \mathscr{L} \\[1mm]
\mc{J}^\mu [\Omega_1, \Omega_2] & = &  (\nabla^\mu \Omega_1) (\nabla_\varphi \Omega_2) + (\nabla^\mu \Omega_2) (\nabla_\varphi \Omega_1)  
+ \delta^\mu_\varphi \mathscr{L}\, ,
\eea
where \hfil $\mathscr{L} = -(\nabla_\mu \Omega_1) (\nabla^\mu \Omega_2) + 2 \Psi_2 \Omega_1\Omega_2$.
\hfil $\mc{E}^\mu$ \hfil and \hfil $\mc{J}^\mu$ \hfil are \hfil conserved \hfil in \hfil the \hfil sense \hfil that\\
$\nabla_\mu \mc{E}^\mu=\nabla_\mu \mc{J}^\mu=0$ if $\Omega_1$
and $\Omega_2$
are both solutions of the F--I equation.

From $\mc{E}^\mu$ and $\mc{J}^\mu$ further conserved currents can be obtained in the following way:
if $\mc{O}$ is a symmetry operator of the F--I equation, then $\mc{E}^\mu [\Omega_1, \mc{O}\Omega_2]$ and $\mc{J}^\mu [\Omega_1, \mc{O}\Omega_2]$
are also conserved currents.
In particular,
\beq
\mc{E}_P^\mu[\Omega] = \mc{E}^\mu [\Omega, P\Omega]
\eeq
and
\beq
\mc{J}_P^\mu[\Omega] = \mc{J}^\mu [\Omega, P\Omega]
\eeq
are conserved if $\Omega$ is a solution of the F--I equation.
$\mc{E}_P^\mu$ and $\mc{J}_P^\mu$ have the properties 
$P\mc{E}_P^\mu = \mc{E}_P^\mu$
and $P\mc{J}_P^\mu = \mc{J}_P^\mu$.
For simplicity (see Section \ref{sec.cctest} and Appendix \ref{app.cl} for further explanation),
we used $\mc{E}_P^\mu$ and $\mc{J}_P^\mu$ to test our code.
A property of $P$ that facilitates its application in our calculations is that it does not change the value of $T$.

We note that, in contrast with the TME, a single F--I equation also follows from a Lagrangian, namely from
$\sqrt{-g}\,  \bigl(-(\nabla_\mu \Omega) (\nabla^\mu \Omega) + 2 \Psi_2 \Omega^2\bigr)$.  

\newpage

\section{Numerical results}
\label{sec.result}

\subsection{Initial data}
\label{sec.initc}

As in \cite{RT, CRT}, we studied the time evolution of field configurations
that have definite azimuthal ($m$) index. In particular, we investigated the cases $m=0,1,2,3$.
(As explained later in this subsection, the cases $m < 0$ are related in a simple way to the cases $m > 0$ by the symmetry $P$,
therefore it is sufficient to consider only non-negative $m$ values.)
We considered `stationary' as well as `nonstationary' initial data (SID and NSID), characterized
by the properties $\partial_T\Psi|_{T=T_0}=0$ and $\Psi|_{T=T_0}=0$, respectively.  
Furthermore, we focused mainly on `pure multipole' initial data, in which only one term in the multipole expansion (see (\ref{eq.ser}))
of $\Psi|_{T=T_0}$ or $\partial_T\Psi|_{T=T_0}$ is nonzero.
The $l$ index of this term is denoted by $l'$. We considered the values $l'=0,1,2,3$.
Pure multipole initial data thus take the form 
\beq
\label{eq.i1}
\Psi|_{T=T_0}=f(R) Y_{l'}^m (\theta,\varphi),\qquad \partial_T\Psi|_{T=T_0}=0
\eeq
or 
\beq
\label{eq.i2}
\partial_T\Psi|_{T=T_0}=f(R) Y_{l'}^m(\theta,\varphi),\qquad \Psi|_{T=T_0}=0.
\eeq
Pure multipole initial data were especially interesting in the case of the TME and the scalar wave equation,
since such data were apparently distinguished, even though the Kerr spacetime does not have $SO(3)$ symmetry.

For $f(R)$ we took the same bump function $\mathscr{B}(R)$ as in \cite{RT,CRT},
which is smooth and has compact support. 
$\mathscr{B}(R)$ is given by the formula
\begin{align}\label{id1}
\mathscr{B}(R)\ &=\ 
\begin{cases}
 \frac{2R}{1-(R^2/M^2)}\exp\left({-\frac{1}{|R-c+w/2|}-\frac{1}{|R-c-w/2|}+\frac{4}{w}}\right)\,,
      & \mathrm{if}\ c-w/2\le R  \le c+w/2,\cr
       \hfill 0\,, &\hfill  \mathrm{otherwise}\,,
\end{cases}
\end{align} 
where $w$ denotes the width of the bump and $c$ determines its center. We considered two different values, $0.7$ and $0.8$, for $c$.
We set the value of $w$ to $0.1$. These values of $c$ and $w$ correspond to a narrow bump not far from the event horizon.
For the Kerr spacetime we mostly chose $a/M=0.5$, and we also considered the case $a=0$.
As mentioned in Section \ref{sec.coords}, we set $M=1$.
For $a/M=0$ and $0.5$, the corresponding value of $R/M$ at the event horizon is approximately $0.41421$ and $0.59864$,
respectively (see also Figure \ref{fig.horizR}).

Although $\psi_{l'}^m$ is real initially, $\psi_l^m$ generally become complex during the time evolution. 
In the case $m=0$, however,
$\Img \psi_l^0$ remains zero if $l-l'$ is even and $\Real \psi_l^0$ remains zero if $l-l'$ is odd, 
as a consequence of the discrete symmetries discussed in Section \ref{sec.discrete}.

From the symmetry of the F--I equation under $P$ and from (\ref{eq.PY}) it also follows that if 
a solution $\Psi$ of the F--I equation has initial data (\ref{eq.i1}) or (\ref{eq.i2}), 
then $P\Psi$ has the same initial data
with $-m$ and multiplied by $(-1)^{l'}$, and the coefficients
in the multipole expansion of $P\Psi$
are related to the coefficients in the multipole expansion of $\Psi$
by the equation $\psi_l^{-m}|_{P\Psi} = (-1)^l(\psi_l^m|_{\Psi})^*$.
Because of this simple relation between the initial data and the solutions with $+m$ and $-m$, 
it is sufficient to investigate only the cases with $m\ge 0$.

If $\Psi$ has a definite azimuthal index, then the expansion (\ref{eq.ser}) can be written as
\beq
\label{eq.ser2}
\Psi(R,T,\theta,\varphi)=\sum_{l=|m|}^\infty \psi_l^m(R,T)Y_l^m(\theta,\varphi),
\eeq
and there are similar expansions for $\Psi_T$ and $\Psi_R$ as well. We kept the first $12$ terms in these series, 
taking into consideration their fast convergence and that the values of $l'$ we chose
in specifying the initial data are not greater than $3$.

In the case of the Schwarzschild spacetime there is no mixing between the different multipole modes,
as has been noted in Section \ref{sec.discrete}, therefore the series (\ref{eq.ser2}) reduces to a single term
\beq
\label{eq.ser2sch}
\Psi(R,T,\theta,\varphi)= \psi_{l'}^m(R,T)Y_{l'}^m(\theta,\varphi)
\eeq
if pure multipole initial data is taken.
From (\ref{eq.de}), one obtains the evolution equation 
\beq
\label{eq.desch}
\partial_{T}^2\psi_{l'}^m  =  \frac{1}{a_{TT}^{(0)}}\left(a_{RR}\,\partial_{R}^2+a_{TR}\,\partial_{TR}
+a_{T}\,\partial_{T}+a_{R}\,\partial_{R}
+a_0 - l'(l'+1)a_{\Delta}
-\frac{2Q^{(0)}}{r}\Psi_2 \right)\,\psi_{l'}^m
\eeq
for $\psi_{l'}^m$. This equation does not depend on $m$.


\subsection{Numerical verification of the conservation of $\mc{E}_P^\mu$ and $\mc{J}_P^\mu$}
\label{sec.cctest}

To test our code, we evaluated numerically
the $Y_0^0$-component of $\sin^2\theta\, \nabla_\mu\mc{E}_P^\mu$ and
$\sin^2\theta\, \nabla_\mu\mc{J}_P^\mu$, which we denote by
$\mc{E}_0^0$ and $\mc{J}_0^0$ ($\mc{E}_P^\mu$ and $\mc{J}_P^\mu$ are introduced in Section \ref{sec.cc}).
We used these divergences instead of the charge balances, that were used in \cite{RT, CRT}, for the sake of novelty,
i.e.\ to explore a slightly different technique of using conservation laws for the verification of numerical computations.
We chose the particular quantities $\mc{E}_0^0$ and $\mc{J}_0^0$ for the relative simplicity of the calculations needed to evaluate them
(see Appendix \ref{app.cl} for additional explanation of the details of the required calculations).
Although restricting one's attention to the $Y_0^0$-component is a minor concession, an analogous concession was also made
in \cite{RT, CRT} by considering only spherically symmetric cylindrical domains in the calculation of charge balances. 
From the transformation properties of the spherical harmonics and of $\mc{E}_P^\mu$ and $\mc{J}_P^\mu$
under $P$ it follows that $\mc{E}_0^0$ and $\mc{J}_0^0$ are real.

\begin{figure}[h]

\vspace{-2cm}
\begin{center}
Conservation law tests
\end{center}
%
%
\includegraphics[scale=0.59]{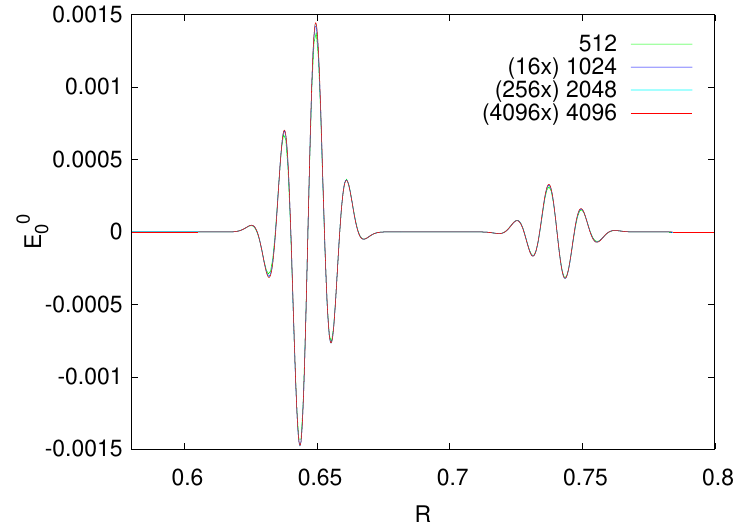} 
\includegraphics[scale=0.59]{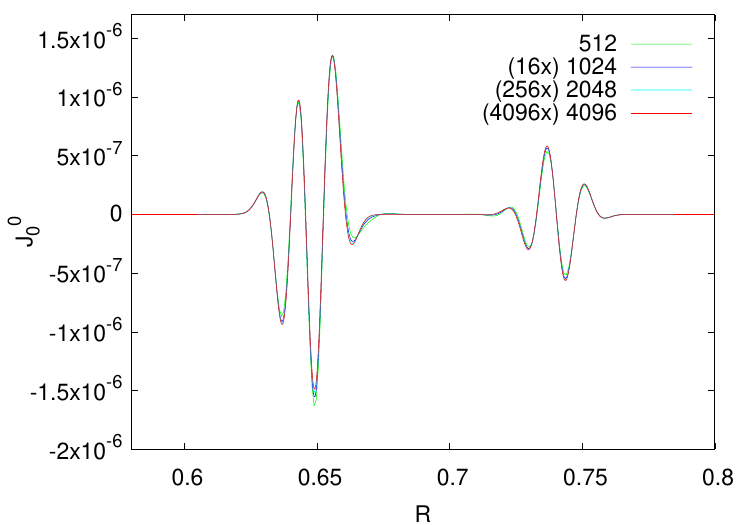} 
\caption{\label{fig.konv1E}
$\mc{E}_0^0$ and $\mc{J}_0^0$ as functions of $R$, at $T\approx 1.477$, for $a=0.5$, $m=1$, $l'=1$, stationary initial data
and grid resolutions $512$, $1028$, $2048$, $4096$, multiplied by $1$, $16$, $256$, $4096$, respectively.
The four lines coincide with very good accuracy, demonstrating that $\mc{E}_0^0$ and $\mc{J}_0^0$ converge to $0$
at the expected rate $\sim (\Delta R)^4 $ as $\Delta R\to 0$.
The number of multipole components taken into account is kept fixed.
}

\vspace{0.3cm}
\includegraphics[scale=0.59]{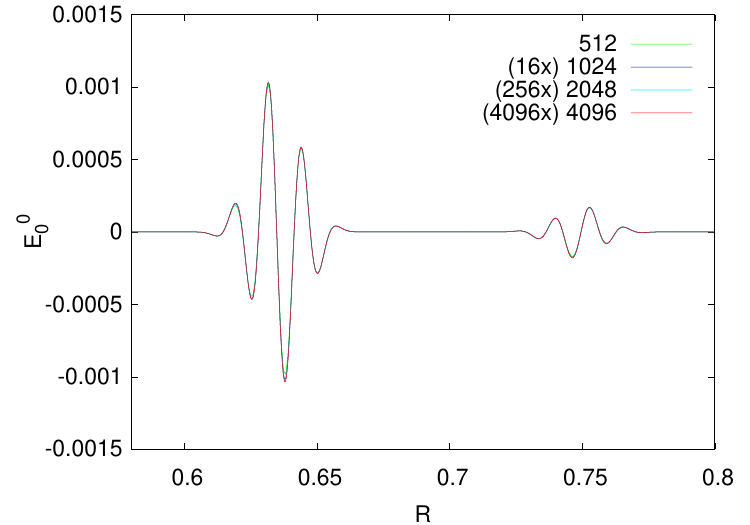} 
\includegraphics[scale=0.59]{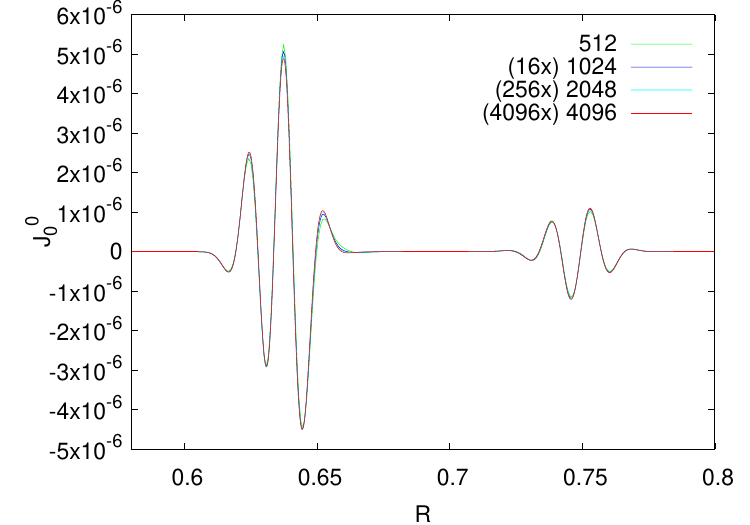} 
\caption{\label{fig.konv2E}
$\mc{E}_0^0$ and $\mc{J}_0^0$ as functions of $R$, at $T\approx 1.64$, for $a=0.5$, $m=-2$, $l'=3$,
grid resolutions $512$, $1028$, $2048$, $4096$, multiplied by $1$, $16$, $256$, $4096$, respectively.
}

\vspace{0.3cm}
\includegraphics[scale=0.59]{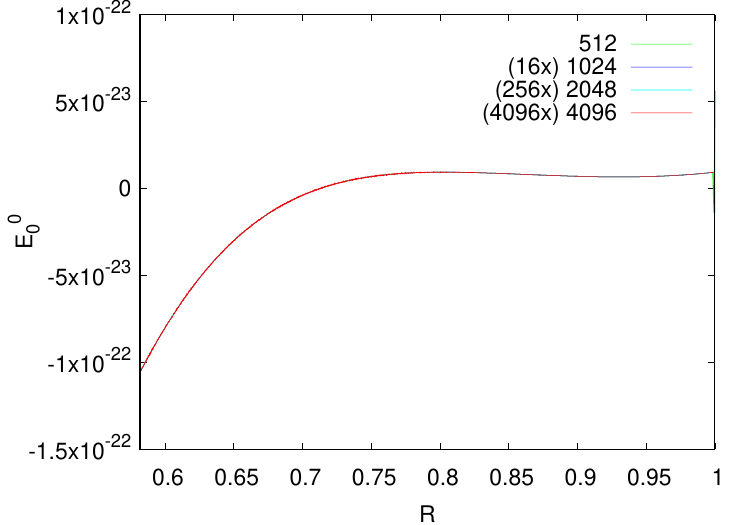} 
\includegraphics[scale=0.59]{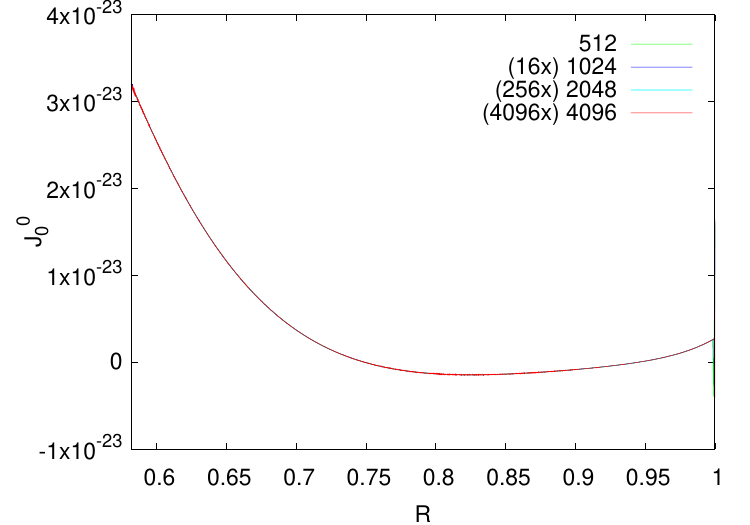} 
\caption{\label{fig.konv2E2}
$\mc{E}_0^0$ and $\mc{J}_0^0$ as functions of $R$, at $T\approx 116.48$, for $a=0.5$, $m=-2$, $l'=3$,
grid resolutions $512$, $1028$, $2048$, $4096$, multiplied by $1$, $16$, $256$, $4096$, respectively.
The lines corresponding to different resolutions lie almost exactly on one another.
The lines for resolution $4096$ are somewhat ragged due to numerical noise, as can be seen by zooming in on them.
}

\end{figure}

If the fourth order Runge--Kutta method is applied, as in the present work, then the numerically computed values of $\mc{E}_0^0$ and $\mc{J}_0^0$
should converge to zero at the rate $\sim(\Delta R)^4$, 
if the grid spacing $\Delta R$ goes to $0$.
We verified this behaviour in several different cases. A few examples are shown in Figures \ref{fig.konv1E}, \ref{fig.konv2E}, \ref{fig.konv2E2}. 
It should be noted that the magnitude of $\mc{E}_0^0$ and $\mc{J}_0^0$ at a single value of $\Delta R$ 
is not of much relevance in itself, 
as it is proportional to the square of the amplitude of $\Omega$.
A remarkable feature of the quantities $\mc{E}_0^0$ and $\mc{J}_0^0$ in comparison with the charge balances used in \cite{RT,CRT},
evident from Figures \ref{fig.konv1E}, \ref{fig.konv2E}, \ref{fig.konv2E2} and from the relevant figures in \cite{RT,CRT},
is that the dependence of $\mc{E}_0^0$ and $\mc{J}_0^0$ on $\Delta R$ follows the law $\sim(\Delta R)^4$ quite accurately,
whereas in the case of the charge balances there are relatively large subleading corrections.

\clearpage

\subsection{Late time behaviour of $\Omega$}
\label{sec.ltd}

In the last phase of the time evolution, $\psi_l^m$, for any $l$, is expected to converge to zero 
according to a power function $\sim T^{-n}$ ($n>0$) as $T\to \infty$, if $m\ne 0$. 
In the case $m=0$, it is possible that $\Psi$ converges to 
$\gamma\Psi_{\mathrm{st}}$ ($\Psi_{\mathrm{st}} = r\Omega_{\mathrm{st}}$; $\Omega_{\mathrm{st}}$ is given by (\ref{eq.omst})),
where $\gamma$ is a constant depending on the initial data,
and then the multipole expansion coefficients of $\Psi - \gamma\Psi_{\mathrm{st}}$ 
are expected to decrease as $\sim T^{-n}$.
If $m\ne 0$, then 
$n$ can be extracted from the numerical results by evaluating the local power indexes (LPIs)
\beq
-\frac{\partial \ln |\Real \psi_l^m|}{\partial \ln T},\qqquad -\frac{\partial \ln |\Img \psi_l^m|}{\partial \ln T}
\eeq
at sufficiently large values of $T$, where they are near their $T\to\infty$ limits.
If $m=0$, then $n$ can be determined from 
\beq
-\frac{\partial \ln |\Real \partial_T \psi_l^m|}{\partial \ln T}-1,\qqquad -\frac{\partial \ln |\Img \partial_T \psi_l^m|}{\partial \ln T}-1,
\eeq
as the differentiation with respect to $T$ eliminates the $\gamma\Psi_{\mathrm{st}}$ part of $\Psi$.

$n$ can take different values at the event horizon, at finite distance from the event horizon
and at $\scrip$; this occurs for instance in the case of the TME.
$n$ can depend on the properties of the initial data
and on the indices of $\psi_l^m$.
The real and imaginary parts of $\psi_l^m$ might also have different decay exponents.

The late time decay exponents were usually found to be integer in previous analytic investigations and
numerical calculations. 
The LPIs we have obtained in the case of the F--I equation are also consistent with
the decay exponents being integer,
therefore we report as decay exponents the integer values that the LPIs appeared to approach as $T$ increased. 

In our investigations we attempted to determine $n$ at several different values of $R$,
namely at the event horizon, at $0.6$, $0.7$, $0.8$, $0.9$, $0.95$, 
and at $\scrip$ ($R=1$).

\subsubsection{Schwarzschild black hole}
\label{sec.nonrot}

\begin{figure}[h]
%
\hspace{-0.5cm}
\includegraphics[scale=0.62]{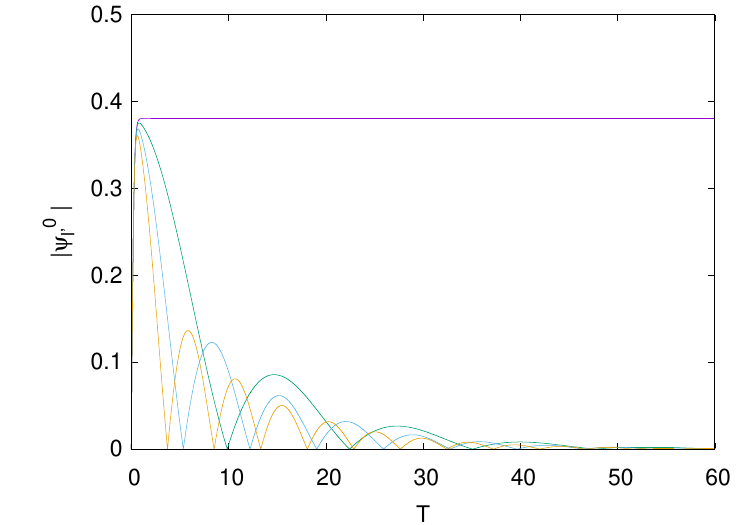} 
\includegraphics[scale=0.62]{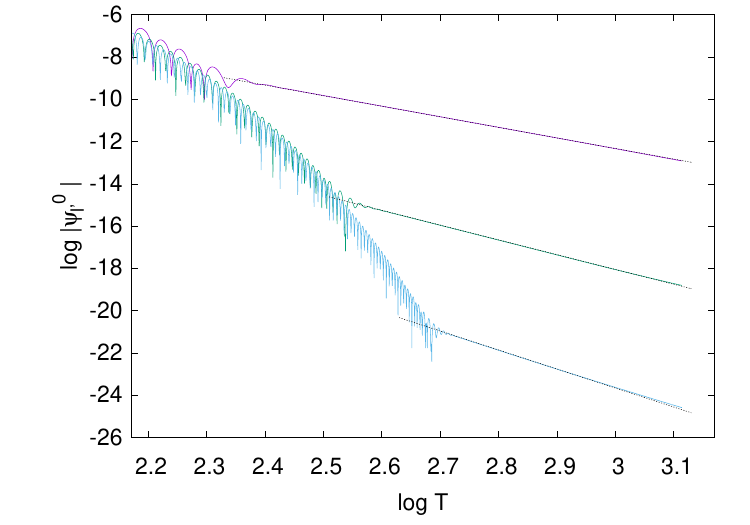}\\ 
\hspace*{3.9cm}(a) \hspace{7.2cm} (b) 
\caption{\label{fig.sch} Plots illustrating the behaviour of $\Psi$ as a function of $T$ in the case $a=0$.\newline
(a) $|\psi_{l'}^0|$ for $l'=0,1,2,3$ at $R=0.7$ with nonstationary initial data;
it can be seen that
$\psi_0^0$ converges to a constant value, whereas $\psi_{l'}^0\to 0$ for $l'=1,2,3$.\newline
(b) $\log |\psi_{l'}^0|$ as a function of $\log T$ for $l'=1,2,3$ at $R=0.7$ with nonstationary initial data,
showing the power-law decay at large $T$.
Dashed black straight lines with slopes $-3$, $-5$ and $-9$ are also plotted for comparison.
The numerical data can be seen to fit well to these lines in the power-law decay phase.}
\end{figure}

In order to investigate the late time behaviour of the solutions of the F--I equation
in the exterior of a Schwarzschild black hole, we carried out numerical simulations
with pure multipole initial data (see Section \ref{sec.initc}) and the methods introduced in the previous sections.
As mentioned in Section \ref{sec.initc}, the evolution equation for $\psi_{l'}^m$ (see (\ref{eq.desch}))
is independent of $m$,
therefore the value of $m$ was not relevant.
The conclusions from our numerical results are the following: 

\begin{enumerate}
\item In agreement with the expectation, 
with spherically symmetric initial data ($m=0$, $l'=0$) $\Psi$ converges to $\gamma\Psi_{\mathrm{st}}$
as $T\to \infty$, whereas with initial data with other values of $m$ or $l'$ it converges to $0$. 
An illustration of this behaviour is provided by Figure \ref{fig.sch}(a). 

\item In the case $m=0$, $l'=0$, the exponential ringdown and the power-law decay tail are completely missing.
This result is explained by the fact that for spherically symmetric $\Omega$ the F--I equation reduces to
$(\partial_t^2-\partial_{r^*}^2)\Omega=0$ (see Section 4.4.\ of \cite{Aksteiner}), where $r^*$ denotes the tortoise coordinate.
For $l'>0$ the exponential ringdown and the power-law decay can be seen in the numerical data;
for an illustration, see Figure \ref{fig.sch}(b). 

\item For any initial data with $l'>0$,
at the event horizon and at finite nonzero distance from the event horizon the decay exponents follow the rule $n=2l'+3$,
whereas at $\scrip$ they can be described by the rule $n=l'+2$.
An illustration of the rule $n=2l'+3$
is provided by Figure \ref{fig.sch}(b),
which shows numerical data at finite distance from the event horizon
together with lines corresponding to exact power functions.
\end{enumerate}

\subsubsection{Rotating black hole}
\label{sec.rot}

\begin{figure}[t]

\hspace{-0.7cm}
\includegraphics[scale=0.62]{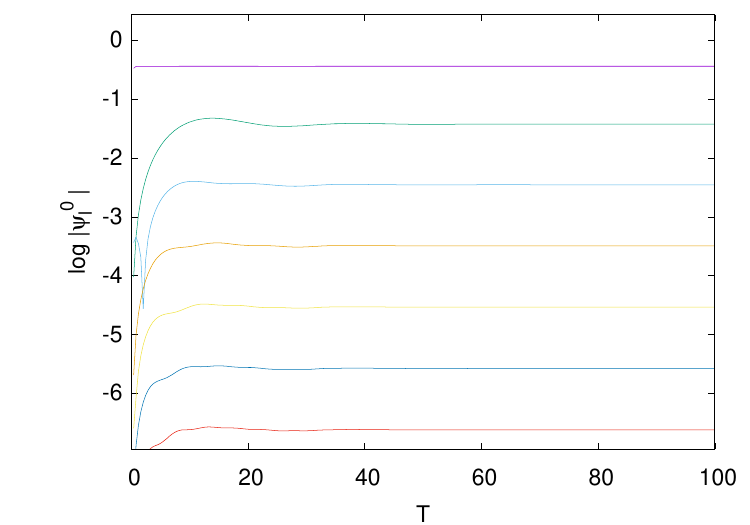} 
\includegraphics[scale=0.62]{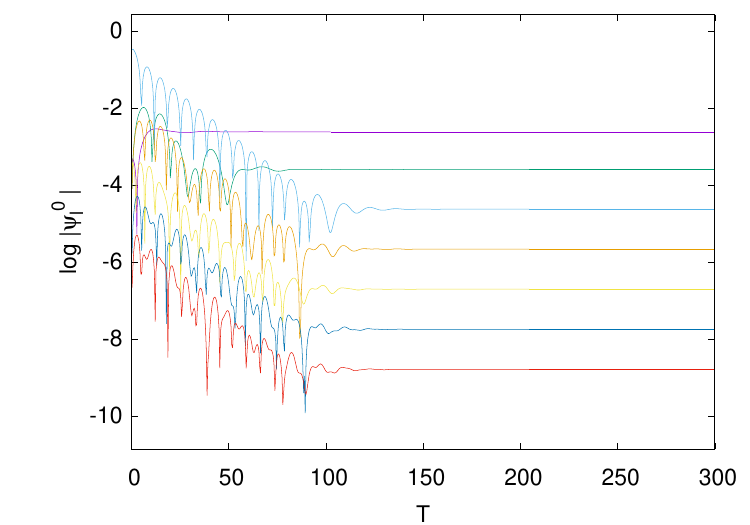}\\ 
\hspace*{3.6cm}(a) \hspace{7.3cm} (b) 
 
\caption{\label{fig.limit1}
Plots illustrating the convergence of $\Psi$ to $\gamma\Psi_{\mathrm{st}}$ as $T\to\infty$
for axially symmetric initial data ($m=0$).\newline 
(a) $\log |\psi_l^0|$ at $R=0.7$ with NSID, $a=0.5$, $l'=0$, $l=0,\dots,6$, $c=0.7$\newline
(b) $\log |\psi_l^0|$ at $R=0.7$ with NSID, $a=0.5$, $l'=2$, $l=0,\dots,6$, $c=0.7$ } 

\end{figure}

\begin{figure}[b]

\hspace{-0.7cm}
\includegraphics[scale=0.62]{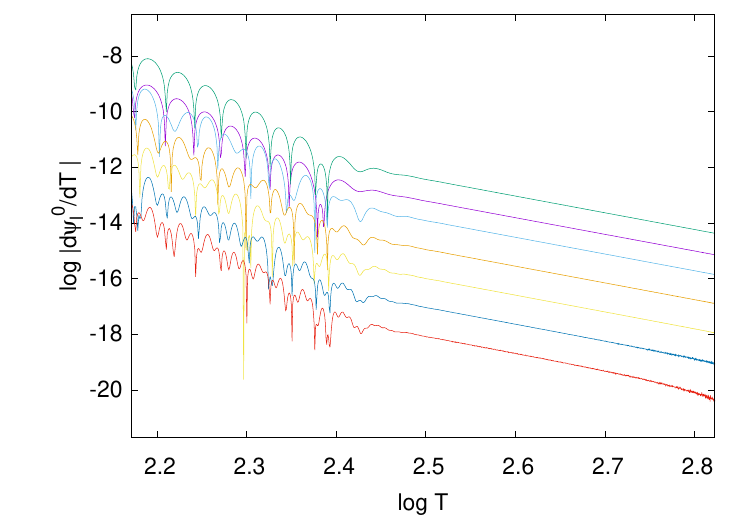}  
\includegraphics[scale=0.62]{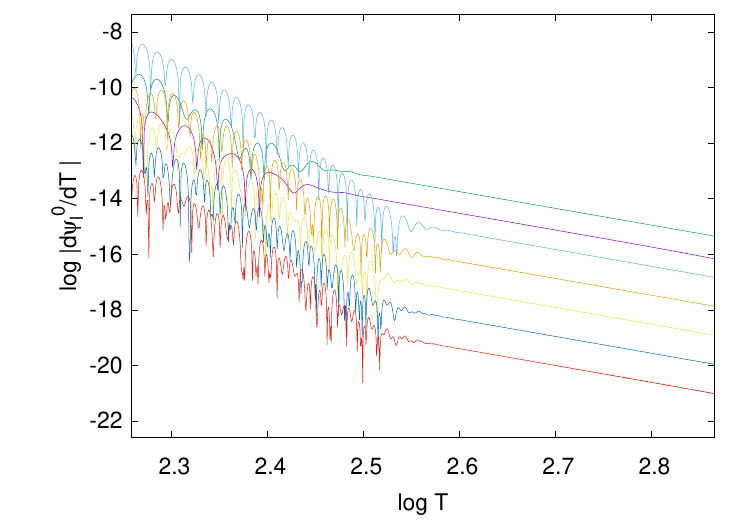}\\ 
\hspace*{3.6cm}(a) \hspace{7.3cm} (b) 
 
\caption{\label{fig.limit1_dT}
Plots illustrating the behaviour of $\partial_T\Psi$ as $T\to\infty$
for axially symmetric initial data ($m=0$). Power-law decay can be seen at large $T$.\newline
(a) $\log |\partial_T\psi_l^0|$ at $R=0.7$ with NSID, $a=0.5$, $l'=0$, $l=0,\dots,6$, $c=0.7$\newline
(b) $\log |\partial_T\psi_l^0|$ at $R=0.7$ with NSID, $a=0.5$, $l'=2$, $l=0,\dots,6$, $c=0.7$ } 

\end{figure}

\begin{figure}

\hspace{-0.7cm}
\includegraphics[scale=0.62]{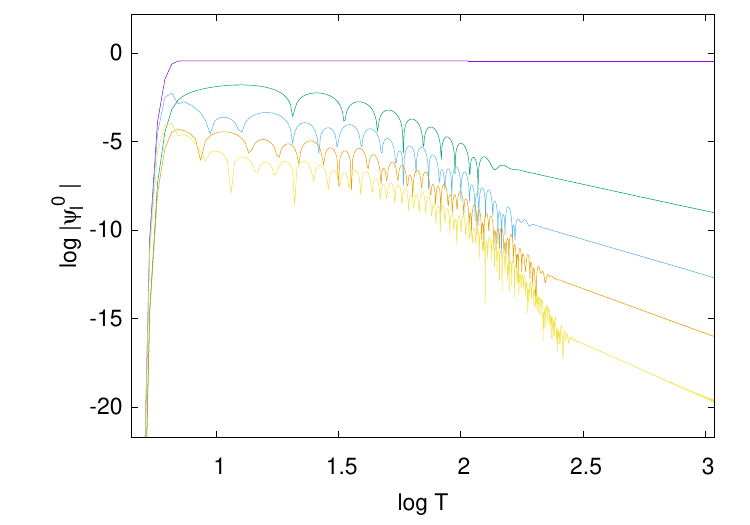} 
\includegraphics[scale=0.62]{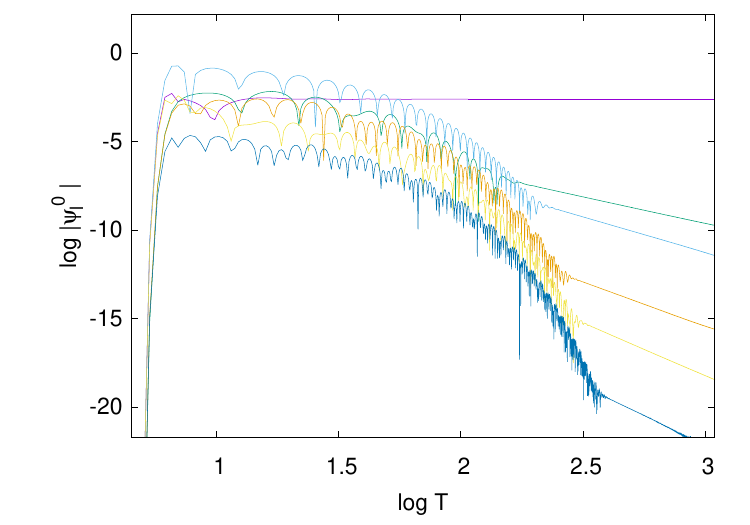}\\ 
\hspace*{3.6cm}(a) \hspace{7.3cm} (b) 
 
\caption{\label{fig.limit2}
Plots illustrating the convergence of $\Psi$ to $\gamma\Psi_{\mathrm{st}}$ as $T\to\infty$ at $\scrip$
for axially symmetric initial data ($m=0$). 
$\Psi_{\mathrm{st}}$ is spherically symmetric at $\scrip$, therefore $\psi_l^0\to 0$ if $l>0$.
The modes with $l>0$ exhibit power-law decay at large $T$. \newline
(a) $\log |\psi_l^0|$ at $R=1$ with NSID, $a=0.5$, $l'=0$, $l=0,\dots,4$, $c=0.7$\newline
(b) $\log |\psi_l^0|$ at $R=1$ with NSID, $a=0.5$, $l'=2$, $l=0,\dots,5$, $c=0.7$ } 

\end{figure}

\begin{figure}

\hspace{-0.3cm}
\includegraphics[scale=0.62]{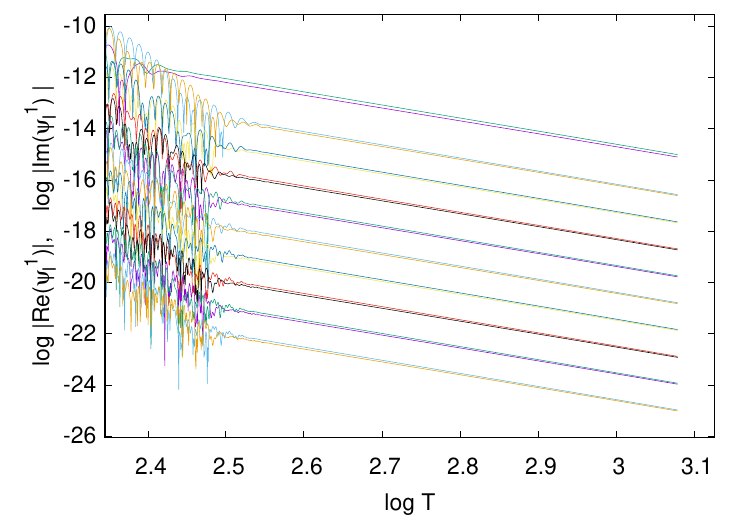}  
\includegraphics[scale=0.62]{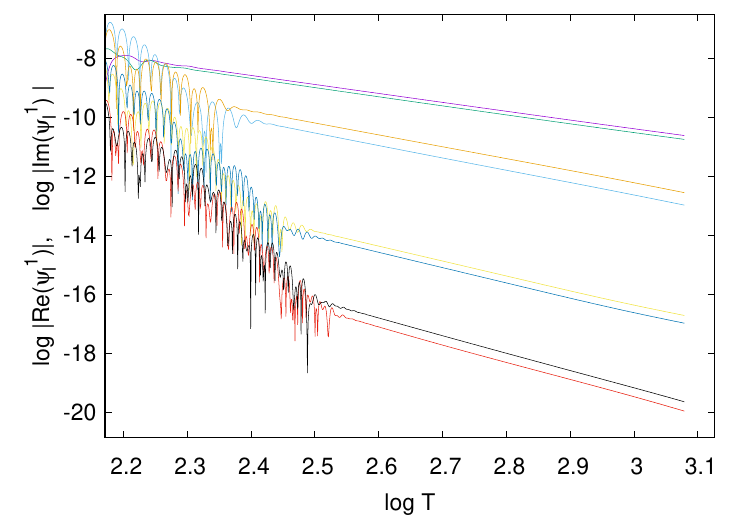}\\ 
\hspace*{3.8cm}(a) \hspace{7.3cm} (b) 
 
\caption{\label{fig.limit3}
Plots illustrating the convergence of $\Psi$ to $0$ as $T\to\infty$ and showing the power-law decay at large $T$,
in the case $m=1$.\newline
(a) $\log |\Real(\psi_l^1)|$ and $\log |\Img(\psi_l^1)|$ as functions of $\log T$ at $R=0.7$ with SID, $a=0.5$, $l'=2$, $l=1,\dots,10$, $c=0.7$\newline
(b) $\log |\Real(\psi_l^1)|$ and $\log |\Img(\psi_l^1)|$ as functions of $\log T$ at $\scrip$
with SID, $a=0.5$, $l'=2$, $l=1,\dots,4$, $c=0.7$} 

\end{figure}

In the investigation of the late time behaviour of the solutions of the F--I equation
in the exterior region of the sub-extremal Kerr spacetime, we set $a=0.5$ in most cases.
We also carried out some calculations at $a=0.1,0.2,\dots,0.9$ to verify the independence of $n$ from $a$;
in accordance with the expectation, we did not find any dependence of $n$ on $a$.
Our main conclusions from the numerical results are the following:

\begin{enumerate}
\item  With axially symmetric initial data (i.e., for $m=0$) $\Psi$ converges to $\gamma\Psi_{\mathrm{st}}$ as $T\to\infty$,
whereas for $m\ne 0$ it converges to $0$, as anticipated.
An illustration of this behaviour is given by Figures \ref{fig.limit1}, \ref{fig.limit1_dT}, \ref{fig.limit2}, \ref{fig.limit3}.
Quasinormal ringdown and power-law decay phases are exhibited by $\psi_l^m$ in all cases
(in the cases with $m=0$, the ringdown and power-law decay phases are present
in the part of $\psi_l^m$ that remains after the subtraction of the static part).
 
\item At $\scrip$ the decay exponents appear to follow the rule $n=l+2$ if $l > 0$.
In the case $m=0$, $l=0$, we found $n=4$.

\item At the event horizon and at finite nonzero distance from the event horizon the decay exponents appear to follow the rule
$n=2|m|+3$ if $m\ne 0$. With axially symmetric initial data (i.e., for $m=0$), we found $n=5$.
The decay exponents at the event horizon do not appear to be different from those at finite nonzero distance from the event horizon.
\end{enumerate}

We note that although generally the rules for the value of $n$ stated above appeared to be valid,
there were some cases when a clear limiting value (as $T\to \infty$) could not be determined from the LPIs.
This occurred, for example, at larger distances from the event horizon
with stationary initial data and the values $m=0$, $l'=2,3$, $l=2$, and $m=0$, $l'=3$, $l=3$.
Nevertheless, the LPIs did not point to smaller values of $n$ than predicted by the above general rules,
i.e.\ the values given by these rules always appeared to be valid at least as lower bounds.
It can be expected that in those cases where the value of $n$ was not clear, the LPI would approach its limit at later times,
which could be reached only with considerably higher numerical accuracy.
The phenomenon that occasionally the LPI approaches its $T\to\infty$ limit only at very late times,
and may first appear to converge to another value, was observed in previous investigations as well \cite{BurKha3,ZenTig,Jasiulek,HBB,RT},
and was thoroughly investigated in \cite{ZenKhaBur}.

A major difference between the decay exponents we have obtained for the F--I equation and
the decay exponents relevant for the TME and the scalar wave equation is that the former show much less dependence on the various parameters.
In particular, in the case of the F--I equation, $n$ does not depend on $l'$ and on whether the initial data is stationary or nonstationary.
Furthermore, $n$ has the same value at the event horizon as at finite nonzero distance from the event horizon,
and this value is also independent of $l$.

Comparing the decay exponents pertaining to the F--I equation and to the scalar wave equation,
at finite distance from the event horizon the decay exponent $|m|+3$ ($|m|>0$) we found equals to the smallest decay exponent
occurring in the case of the scalar wave equation at the same value of $m$.
At $\scrip$, the smallest decay exponent occurring in the case of the scalar wave equation at a given value of $l$ is $l+2$;
for $l>0$, this agrees with the decay exponents we found in the case of the F--I equation.

In contrast with our experience with the TME and the scalar wave equation \cite{RT,CRT},
at finite distance from the event horizon a clear value of $n$ could be seen up to relatively high values of $l$
(up to $l\simeq 10$ at least) in many cases; Figure \ref{fig.limit3}(a), for instance, shows such a case.
This is probably related to the fact that $n$ does not increase with $l$ for fixed initial data;
numerical errors can be expected to tend to swamp the LPIs of $\psi_l^m$ with higher values of $l$ if the corresponding
decay exponents increase with $l$.
At $\scrip$, a similar difference between the F--I equation and the TME cannot be seen;
we could determine the value of $n$ only up to $l=4$ or $5$ at $\scrip$.

\newpage

\newpage

\clearpage

\section{Conclusion}
\label{sec.concl}

We investigated the late-time behaviour of the solutions of the Fackerell--Ipser equation numerically
on the domain of outer communication of sub-extremal Kerr spacetime, including the event horizon and future null infinity.
We considered the case of the Schwarzschild background as well.
We used largely the same methods as in our previous investigations of the late-time tails of the solutions of the scalar wave equation
and the Teukolsky master equation \cite{RT,CRT}, but some technical novelties also appear in the present work.
Instead of charge balances, we computed divergences of conserved currents
to test our code for solving the Fackerell--Ipser equation with given initial data.
This required some additional techniques (described in Appendix \ref{app.cl})
regarding the algebraic treatment of the expansion of the field variable
in terms of spherical harmonic functions.
In the construction of the conserved currents
we took into account a discrete symmetry of the Fackerell--Ipser equation (see (\ref{eq.P})).
This symmetry was also useful for reducing the number of different cases that had to be investigated numerically.
Since the knowledge of the radial incoming and outgoing speed of light is important for determining
the optimal ratio of time step to radial grid spacing,
and we had not written about them in detail in our previous works,
we included an appendix where they are examined in our coordinate system (see Appendix \ref{app.lightspeed}).

For the initial data, we took pure multipole configurations with compact support
and being either stationary or non-stationary. As in our previous investigations,
we considered both axisymmetric and non-axisymmetric configurations.
We found that with such initial data the solutions of the Fackerell--Ipser equation converge at late times
either to a known static solution (up to a constant factor) or to zero. This result is in agreement with \cite{AndBlu}.
Note, however, that while the result of \cite{AndBlu} was obtained for very slowly rotating black holes,
in the present investigation the value of the rotation parameter $a/M$ of the Kerr spacetime was not small
(it was $0.5$ in most of the calculations).

As the solutions approached the late-time limit, they exhibited a quasinormal ringdown and finally a power-law decay.
We extracted the exponents characterizing the power-law decay of the spherical harmonic components of the field variable
from the numerical data for various values of the parameters of the initial data,
and based on the results we made a proposal for Price's law for the Fackerell--Ipser equation in sub-extremal Kerr spacetime
and Schwarzschild spacetime
(see Sections \ref{sec.nonrot} and \ref{sec.rot}).

In comparison with the Teukolsky master equation and the scalar wave equation, a remarkable feature of the decay exponents
in the case of the Fackerell--Ipser equation (in rotating background)
is that they show much less dependence on the parameters pertaining to the initial data
and the late-time field.
This behaviour might be related to the nonintegrability of the Fackerell--Ipser equation,
however, further investigation would be needed to determine if such a relation exists.

In the case of the Schwarzschild background,
we found that generally the decay exponents are the same as for the scalar wave equation,
except in the case of the spherically symmetric configurations,
where the power-law late-time tails and the quasinormal ringdown are completely missing.

It would be interesting to extend the present investigations to near-extremal and extremal Kerr backgrounds.
In view of the results of previous studies
\cite{AndGla1,AndGla2,HBB,YanZimZen,CasGraZim,GraZim,BurKha5,CasZim},
the late-time behaviour of the solutions of the Fackerell--Ipser equation
on extremal Kerr background can be expected to be different from
their behaviour on non-extremal Kerr backgrounds.
However, the methods would need to be modified somewhat,
because, as explained in Section \ref{sec.numm},
in the extremal and very close to extremal backgrounds
the expansion (\ref{eq.11x}) cannot be used to do the division by the denominator of $\Psi_2$
(see (\ref{eq.psi2}) for this denominator).

It would also be of some interest to observe numerically the transition from rotating to non-rotating background.
A further possible extension of the present study would be to consider initial data that fall off slowly toward future null infinity
(for example, according to power functions), similarly to \cite{RT}.
Investigating the late-time tails of the solutions of the Fackerell--Ipser equation using analytic methods
would also be interesting.

\appendix

\renewcommand{\theequation}{\thesection.\arabic{equation}}

\section{Coefficients of the Fackerell--Ipser equation}
\label{app.coeff}

The coefficients $a_{RR}$, $a_{TR}$, $a_{T\varphi}$, $a_{R\varphi}$, $a_T$, $a_R$, $a_\varphi$, $a_0$, $a_\Delta$,
$a_{TT}^{(0)}$, $a_{TT}^{(2)}$, $Q^{(0)}$ and $Q^{(2)}$ 
appearing in equation (\ref{eq.de}) are the following:
{
\small
\bea
a_{RR} & = &
-\frac{(M-R)^2(M+R)^5 (4 M^3 R (M^2-M R-R^2)-a^2 (M^2-R^2)^2)}{4 M^6 R
   (M^2+R^2)^2} \\
a_{TR} & = & 
\frac{2 (M+R)^3}{M^5 (M^2+R^2)^2} \bigl(2 M^3 (M^5+6 M^4 R-6 M^3 R^2-10 M^2 R^3+3 M R^4+4 R^5) \nonumber \\ 
&& -a^2 (3 M^2-2 R^2) (M^2-R^2)^2\bigr) \\
a_{T\varphi} & = & -\frac{4a(M+R)^3(3M^2-2R^2)}{M^3(M^2+R^2)}  \\
a_{R\varphi} & = & \frac{a(M-R)^2(M+R)^5}{M^4R(M^2+R^2)}    \\
a_{T} & = &
-\frac{(M-R) (M+R)^3}{M^5 R (M^2+R^2)^3}
\bigl(a^2 (-3 M^7-3 M^6 R-20 M^5 R^2-20 M^4 R^3+9 M^3 R^4 \nonumber \\ 
&& +9 M^2 R^5+6 M R^6+6 R^7)+2 M^3
   (M^6+M^5 R+10 M^4 R^2+42 M^3 R^3 \nonumber \\ 
&&   +27 M^2 R^4+11 M R^5+8 R^6)\bigr)
\\
a_{R} & = &
\frac{(M-R)(M+R)^4 }{2 M^6 R^2 (M^2+R^2)^3} \bigl(2 M^3 R (M^6+7 M^4 R^2-6 M^3 R^3-5 M^2 R^4-2 M R^5-3 R^6) \nonumber \\
&& -a^2 (M^2-R^2)^2
   (M^4+5 M^2 R^2+2 R^4)\bigr)
\\
a_{\varphi} & = & -\frac{a(M-R)(M+R)^4}{M^4R^2}  \\
a_{0} & = & -\frac{(M-R)(M+R)^4}{2M^6R^3}\bigl(2M^3R+a^2(R^2-M^2)\bigr)
\\
a_\Delta & = & 
\frac{(M+R)^3}{M^2R}
\\
a_{TT}^{(0)} & = & 
\frac{ (M+R)^2}{15 M^4 R (M^2+R^2)^2}
\bigl(5 a^2 (M+R) (M^6-106 M^4 R^2+145 M^2 R^4-48 R^6) \nonumber\\
&& +60 M^3 R (M^5+14 M^4 R+52 M^3
   R^2+23 M^2 R^3-24 M R^4-16 R^5)\bigr) \\
a_{TT}^{(2)} & = & \frac{4}{3}\sqrt{\frac{\pi}{5}}\frac{a^2(M+R)^3}{M^2R}\\
Q^{(0)} & = & \frac{1}{(M-R)^3}\left( 8 M^4 R^2 + \frac{2}{3} a^2 (M^2-R^2)^2 \right)\\
Q^{(2)} & = & \frac{8}{3}\sqrt{\frac{\pi}{5}} a^2 \frac{(M+R)^2}{M-R} 
\eea

}

\newpage

\section{Radial light speeds}
\label{app.lightspeed}

In this appendix the ingoing and outgoing radial light speeds in the $(T,R,\theta,\varphi)$ coordinates are discussed.
These light speeds, denoted by $c_\pm$, are defined by the equations
\beq
\label{eq.null}
v^\mu = (\partial_T)^\mu + c_{\pm}(\partial_R)^\mu\, ,\qquad v_\mu v^\mu=0,
\eeq
i.e.\ $c_\pm$ are the velocities of the null curves lying in the $\theta,\varphi = const$ planes. 
From (\ref{eq.null}) one obtains the formula
\beq
\label{eq.lspeed}
c_{\pm}=\frac{\pm \sqrt{g_{TR}^2-g_{TT}g_{RR}}-g_{TR}}{g_{RR}}.
\eeq
$g_{TR}^2-g_{TT}g_{RR}$ turns out to be a complete square; $\sqrt{g_{TR}^2-g_{TT}g_{RR}}=2M^2(M^2+R^2)/(M^2-R^2)^2$.
The ingoing speed $c_+$ is independent of $a$ and $\theta$ and has a relatively simple expression in terms of $M$ and $R$:
\beq
c_+ = -\frac{(M+R)(M-R)^2}{2M(M^2+5MR-4R^2)}.
\eeq
For the outgoing speed $c_-$ one obtains a more complicated formula:
\bea
c_- & = &  -\bigl((M + R)^2 (-a^2 (M^2 - R^2)^2 + 8 M^3 R (M^2 - M R - R^2) - a^2 (M^2 - R^2)^2 \cos(2\theta))\bigr)/  \nonumber \\
&& \hspace{1cm} /\bigl(2 M (a^2 (M + R)^2 (M^3 - 6 M^2 R + M R^2 + 4 R^3)  \nonumber  \\
&& \hspace{3cm} +  8 M^3 R (M^3 + 9 M^2 R + 11 M R^2 + 4 R^3)  \nonumber \\
&& \hspace{3cm} +  a^2 (M + R)^2 (M^3 - 6 M^2 R + M R^2 + 4 R^3) \cos(2\theta))\bigr).
\eea
In the Schwarzschild limit $c_-$ simplifies to
\beq
c_-|_{a=0}\ =\ -\frac{(M+R)^2(M^2-MR-R^2)}{2M(M^3+9M^2R+11MR^2+4R^3)}.
\eeq
Moreover, $c_-$ is independent of $a$ at $\theta=\pi/2$, so
\beq
c_-|_{\theta=\pi/2} \ = \ c_-|_{a=0}\, .
\eeq
We note that $g_{TT}$, $g_{RR}$ and $g_{TR}$ are also independent of $a$ at $\theta=\pi/2$.

$c_+$ is a monotonic increasing function of $R/M$; it reaches zero at $\scrip$ (see Figure \ref{fig.cpm05}(a)).
The
value of $c_+$ at the event horizon is
$\approx\ -0.0460655,\ -0.0503023,\ -0.0558808,\ -0.0689487,\ -0.101746$ at
$a/M\, =\, 0,\ 0.5,\ 0.7,\ 0.9,\ 1.0$, respectively.

\begin{figure}[h]
%
\begin{center}
\hspace*{1cm}$a/M=0.7$\hspace{6cm}$a/M=0.9$   \\
\includegraphics[height=4.8cm]{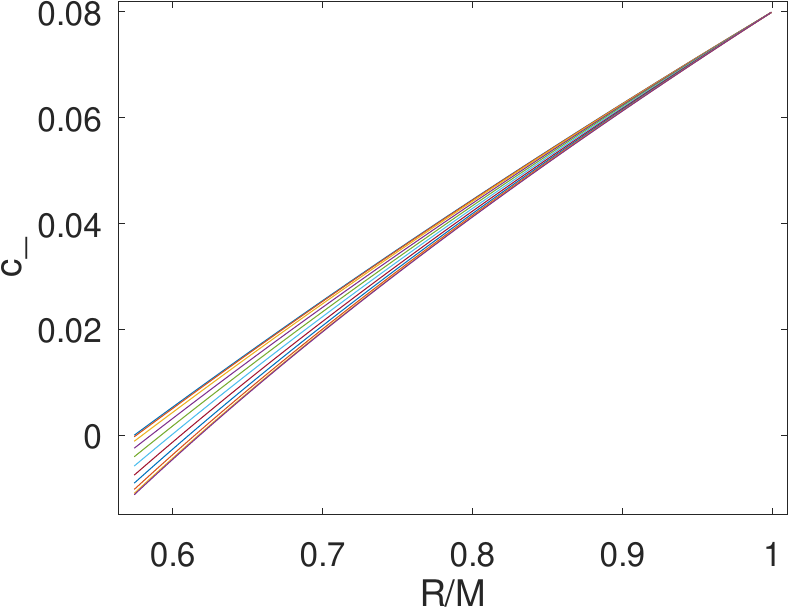} 
\hspace{1cm}
\includegraphics[height=4.8cm]{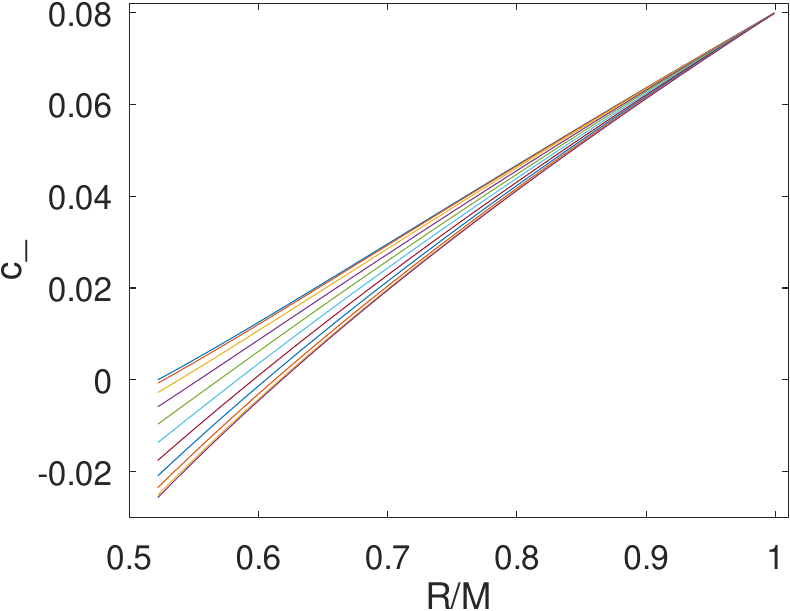}\\[4mm] 
\hspace*{1cm}$a/M=0$\hspace{6cm}$a/M=1$   \\
\includegraphics[height=4.8cm]{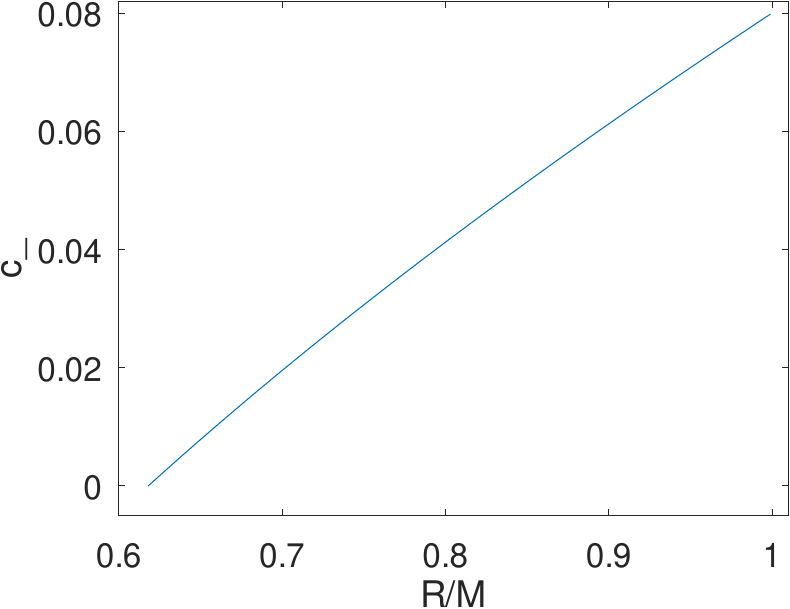} 
\hspace{1cm}
\includegraphics[height=4.8cm]{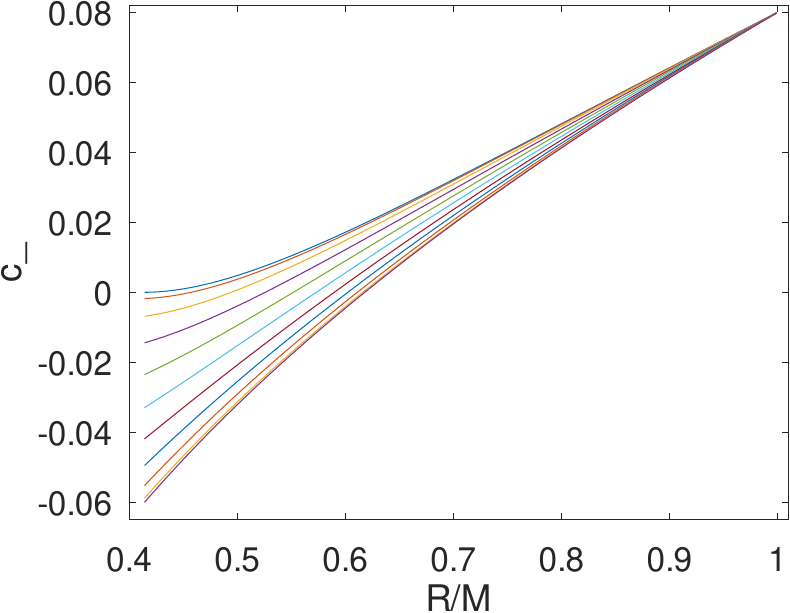} 
\end{center}
\vspace{-0.5cm}
\caption{\label{fig.cm07-09} The outgoing speed of light $c_-$ as a function of $R/M$ in the range $[R_+/M,1]$ at $a/M=0.7, 0.9, 0, 1$
and $\theta=(\pi/2)(n/10)$, $n=0,1,\dots, 10$}
\end{figure}

It can be seen by inspection that $c_-$ is also a monotonic increasing function of $R/M$ at any fixed $\theta$
(see Figure \ref{fig.cpm05}(b) and \ref{fig.cm07-09} for plots of $c_-$).
Its maximal value, reached at $\scrip$, is $0.08$, independently of the value of $a$ and $\theta$.
$c_-$ is zero at the ergosphere, where $g_{TT}=0$, i.e.\ $(\partial_T)^\mu$ becomes null
($(\partial_T)^\mu$ is timelike outside the ergosphere and spacelike between the ergosphere and the event horizon).
At fixed $R/M$, $c_-$ increases as $\theta$ varies from $\pi/2$ to $0$ or $\pi$.

The main conclusion from the above considerations is that
the maximum of $|c_-|$ and $|c_+|$ in the exterior region is not greater than $0.08$ if $a/M \le 0.9$.
$|c_+|$ is maximal at the event horizon, whereas $|c_-|$ is maximal at $\scrip$.
The maximum of $|c_-|$ is $0.08$ for all values of $a/M$.

By checking with solutions of the F--I equation
generated numerically from initial data with compact support,
it can be seen that the speed of the propagation of the inner and outer boundary of the support of the coefficients $\psi_l^m(R,T)$
is given by $c_+$ and $c_-|_{\theta=0}$, respectively.

\clearpage

\section{Technical details of the verification of the conservation laws}
\label{app.cl}

In this appendix we discuss some of the technical details of the numerical calculations needed in the 
verification of the conservation laws mentioned in Section \ref{sec.cc}.

For calculating $\nabla_\mu \mc{E}_P^\mu$ and $\nabla_\mu \mc{J}_P^\mu$, a possible approach is to use the formula
\beq
\label{eq.id-nablav}
\nabla_\mu v^\mu = (\sqrt{-g})^{-1}\partial_\mu (\sqrt{-g}v^\mu).
\eeq
In the present case,
\beq
\sqrt{-g} = F(R,\theta) \sin\theta\, ,\qquad
F=\frac{M^2(M^2+R^2)}{(M-R)^4(M+R)^4}(M-R)^3 Q\, .
\label{eq.sqrtg}
\eeq
However, from the point of view of spherical harmonics, the techniques mentioned in Section \ref{sec.method}
are not sufficient for evaluating $\nabla_\mu \mc{E}_P^\mu$ and $\nabla_\mu \mc{J}_P^\mu$. In fact, the calculation of 
$\nabla_\mu \mc{E}_P^\mu$ and $\nabla_\mu \mc{J}_P^\mu$ appears to be very difficult, as it would involve multiplication and division
of spherical harmonics by $\sin\theta$, and differentiation of spherical harmonics with respect to $\theta$.  
In order to avoid these problems, we consider $\sin^2\theta\,\nabla_\mu \mc{E}_P^\mu$ and $\sin^2\theta\,\nabla_\mu \mc{J}_P^\mu$
instead of $\nabla_\mu \mc{E}_P^\mu$ and $\nabla_\mu \mc{J}_P^\mu$,
since it turns out that the calculation of these quantities is manageable. 
In addition to the techniques mentioned in Section \ref{sec.method},
it requires only one more kind of operation, 
which is the evaluation of the action of the differential operator $\sin\theta\,\partial_\theta$ on spherical harmonics.
As explained in the next paragraph, this does not present much difficulty.

For the calculation of $\sin\theta\, \partial_\theta Y_l^m$, the formula
\beq
\label{eq.sthdth}
\sin\theta\,\partial_\theta \Phi = \sqrt{\frac{2\pi}{3}}\, (Y_1^{-1}J_+ \Phi + Y_1^1 J_- \Phi)
\eeq
can be used,
where $\Phi(\theta,\varphi)$ is an arbitrary function defined on the sphere,
$Y_1^{-1}$ and $Y_1^1$ are the spherical harmonics
\beq
Y_1^{-1}=\frac{1}{2}\sqrt{\frac{3}{2\pi}}\sin\theta\, \ee^{-\ii\varphi}\, , \qqquad
Y_1^{1}=-\frac{1}{2}\sqrt{\frac{3}{2\pi}}\sin\theta\, \ee^{\ii\varphi}\, ,
\eeq
and the operators $J_\pm$ are the standard $\mathfrak{so}(3)$ Lie algebra elements
that raise and lower the $z$-component of the spin.
$J_\pm$ act generally as
\beq
J_+ \Phi = \biggl(\ee^{\ii\varphi}\partial_\theta + \ii \frac{\cos\theta}{\sin\theta}\ee^{\ii\varphi}\partial_\varphi\biggr)\Phi\, ,\qquad
J_- \Phi = \biggl(-\ee^{-\ii\varphi}\partial_\theta + \ii \frac{\cos\theta}{\sin\theta}\ee^{-\ii\varphi}\partial_\varphi\biggr)\Phi\, ,
\eeq
and their action on spherical harmonics is
\beq
J_+Y_l^m = \sqrt{(l+m+1)(l-m)}\,Y_l^{m+1}\, , \qquad
J_-Y_l^m = \sqrt{(l-m+1)(l+m)}\,Y_l^{m-1}\, .
\eeq
The latter formulae can be used to evaluate $J_+\Phi$ and $J_-\Phi$ on the right hand side of (\ref{eq.sthdth})
for $\Phi=Y_l^m$, and then the multiplications by $Y_1^{-1}$ and $Y_1^1$ can be executed in the 
usual way, as described in Section \ref{sec.method}.

The components of $g^{\mu\nu}$ are also needed 
for calculating $\sin^2\theta\,\nabla_\mu \mc{E}_P^\mu$ and $\sin^2\theta\,\nabla_\mu \mc{J}_P^\mu$. 
They are given by the following expressions:
{\small
\bea
g^{TT} & = &\frac{2M^2(M-R)^2 R}{(M+R)}\frac{a_{TT}}{\tilde{Q}} \\
g^{TR} & = &
-\frac{1}{\tilde{Q}}\frac{2 R (M-R)^2 (M+R)^2}{M^3 (M^2+R^2)^2}
\bigl(2 M^3 (M^5+6 M^4 R-6 M^3 R^2 \nonumber\\
&& -10 M^2 R^3+3 M R^4+4 R^5)-a^2 (3 M^2-2 R^2)
(M^2-R^2)^2\bigr)\\
g^{T\varphi} & = & \frac{4 a R (M-R)^2 (M+R)^2 (3 M^2-2 R^2)}{M (M^2+R^2)}\frac{1}{\tilde{Q}}\\
g^{RR} & = &
\frac{(M-R)^4 (M+R)^4 (4 M^3 R (M^2-M R-R^2)-a^2 (M^2-R^2)^2)}{2 M^4 (M^2+R^2)^2}\frac{1}{\tilde{Q}}\\
g^{R\varphi} & = & -\frac{a(M-R)^4(M+R)^4}{M^2(M^2+R^2)}\frac{1}{\tilde{Q}}\\
g^{\theta\theta} & = & -2(M-R)^2(M+R)^2\frac{1}{\tilde{Q}}\\
g^{\varphi\varphi} & = & -2(M-R)^2(M+R)^2\frac{1}{\tilde{Q}\sin^2\theta}\, ,
\eea
}
where $\tilde{Q}=(M-R)^3 Q$.

$\sqrt{-g}$ is singular at $R=M$ due to the factor $(M-R)^4$ in the denominator of $F$ (see (\ref{eq.sqrtg})),
therefore it would seem, taking into consideration the identity (\ref{eq.id-nablav}),
that $\sin^2\theta\,\nabla_\mu \mc{E}_P^\mu$ and $\sin^2\theta\,\nabla_\mu \mc{J}_P^\mu$
are also singular.
Nevertheless, the singularity is cancelled due to the presence of factors of the type $(M-R)^n$ ($n$ being a positive integer)
in $g^{RR}$, $g^{RT}$, $g^{R\varphi}$.
This cancellation can be taken into account in the numerical calculations,
i.e.\ quantities that are singular at $R=M$ can be avoided.

For the evaluation of $\mc{E}_0^0$ and $\mc{J}_0^0$ it is necessary to calculate the $Y_0^0$-component of products
$W_1W_2$, where $W_1$ and $W_2$ are sums of the form
$W_1=\sum_{l} w_{1l}^m Y_l^m$, $W_2=\sum_{l} w_{2l}^{-m}Y_l^{-m}$, with known coefficients $w_{1l}^m$, $w_{2l}^{-m}$.
The $Y_0^0$-component of such products
is given by the simple formula $(-1)^m \frac{1}{\sqrt{4\pi}} Y_0^0 \sum_{l} w_{1l}^m w_{2l}^{-m}$.
The simplicity of the latter formula is one of the main reasons we use
$\mc{E}_P^\mu$ and $\mc{J}_P^\mu$, and in particular $\mc{E}_0^0$ and $\mc{J}_0^0$, for testing our code.

\section*{Acknowledgments}

The authors are indebted to Lars Andersson for suggesting the problem and for useful guidance.
We also thank Steffen Aksteiner and Piotr Bizon for helpful discussions. This project was supported in part by NKFIH grant K-142423.


\small

\begin{thebibliography}{99}

\bibitem{FI}Fackerell, E. D., Ipser, J. R.:
Weak electromagnetic fields around a rotating black hole.
Phys. Rev. D {\bf 5} 2455 (1972).
\href{https://doi.org/10.1103/PhysRevD.5.2455}{https://doi.org/10.1103/PhysRevD.5.2455}

\bibitem{Teukolsky1}Teukolsky, S. A.:
Rotating black holes: separable wave equations for gravitational and electromagnetic perturbations.
Phys. Rev. Lett. {\bf 29} 1114 (1972).
\href{https://doi.org/10.1103/PhysRevLett.29.1114}{https://doi.org/10.1103/PhysRevLett.29.1114}

\bibitem{Teukolsky2}Teukolsky, S. A.:
Perturbations of a rotating black hole. I. Fundamental equations for gravitational, electromagnetic, and neutrino-field perturbations.
Astrophys. J. {\bf 185} 635-647 (1973).

\bibitem{CasOrt}Torres del Castillo, G. F., Silva-Ortigoza, G.:
Spin-3/2 perturbations of the Kerr-Newman solution.
Phys. Rev. D {\bf 46} 5395-5398 (1992).
\href{https://doi.org/10.1103/PhysRevD.46.5395}{https://doi.org/10.1103/PhysRevD.46.5395}

\bibitem{Ort}Silva-Ortigoza, G.:
Killing spinors and separability of Rarita-Schwinger's equation in type
\{2,2\} backgrounds.
J. Math. Phys. {\bf 36} 6929-6936 (1995).
\href{https://doi.org/10.1063/1.531199}{https://doi.org/10.1063/1.531199}


\bibitem{Price1}Price, R. H.:
Nonspherical perturbations of relativistic gravitational collapse. I. Scalar and gravitational perturbations.
Phys. Rev. D \textbf{5} 2419 (1972).
\href{https://doi.org/10.1103/PhysRevD.5.2419}{https://doi.org/10.1103/PhysRevD.5.2419}

\bibitem{Price2}Price, R. H.:
Nonspherical perturbations of relativistic gravitational collapse. II. Integer-spin,
zero-rest-mass fields.
Phys. Rev. D {\bf 5} 2439 (1972).
\href{https://doi.org/10.1103/PhysRevD.5.2439}{https://doi.org/10.1103/PhysRevD.5.2439}



\bibitem{Whiting}Whiting, B. F.:
Mode stability of the Kerr black hole.
J. Math. Phys. {\bf 30} 1301 (1989).
\href{https://doi.org/10.1063/1.528308}{https://doi.org/10.1063/1.528308}

\bibitem{KriLagPap}Krivan, W., Laguna, P., Papadopoulos, P.:
Dynamics of scalar fields in the background of rotating black holes.
Phys. Rev. D \textbf{54} 4728 (1996).
\href{https://doi.org/10.1103/PhysRevD.54.4728}{https://doi.org/10.1103/PhysRevD.54.4728}

\bibitem{KriLagPapAnd}Krivan, W., Laguna, P., Papadopoulos, P., Andersson, N.:
Dynamics of perturbations of rotating black holes.
Phys. Rev. D {\bf 56} 3395-3404 (1997).
\href{https://doi.org/10.1103/PhysRevD.56.3395}{https://doi.org/10.1103/PhysRevD.56.3395}

\bibitem{Krivan}Krivan, W.:
Late-time dynamics of scalar fields on rotating black hole backgrounds.
Phys. Rev. D \textbf{60} 101501(R) (1999).
\href{https://doi.org/10.1103/PhysRevD.60.101501}{https://doi.org/10.1103/PhysRevD.60.101501}

\bibitem{BarOri}Barack, L., Ori, A.:
Late time decay of scalar perturbations outside rotating black holes.
Phys. Rev. Lett. \textbf{82} 4388 (1999).
\href{https://doi.org/10.1103/PhysRevLett.82.4388}{https://doi.org/10.1103/PhysRevLett.82.4388}

\bibitem{Barack}Barack, L.:
Late time decay of scalar, electromagnetic, and gravitational perturbations outside rotating black holes.
Phys. Rev. D \textbf{61} 024026 (1999).
\href{https://doi.org/10.1103/PhysRevD.61.024026}{https://doi.org/10.1103/PhysRevD.61.024026}

\bibitem{BarOri2}Barack, L., Ori, A.:
Late time decay of gravitational and electromagnetic perturbations along the event horizon.
Phys. Rev. D \textbf{60} 124005 (1999).
\href{https://doi.org/10.1103/PhysRevD.60.124005}{https://doi.org/10.1103/PhysRevD.60.124005}

\bibitem{Hod1}Hod, S.:
Mode coupling in rotating gravitational collapse of a scalar field.
Phys. Rev. D \textbf{61} 024033 (1999).
\href{https://doi.org/10.1103/PhysRevD.61.024033}{https://doi.org/10.1103/PhysRevD.61.024033}

\bibitem{Hod2}Hod, S.:
Radiative tail of realistic rotating gravitational collapse.
Phys. Rev. Lett. \textbf{84} 10-13 (2000).
\href{https://doi.org/10.1103/PhysRevLett.84.10}{https://doi.org/10.1103/PhysRevLett.84.10}

\bibitem{Hod3}Hod, S.:
Mode coupling in rotating gravitational collapse: Gravitational and electromagnetic perturbations.
Phys. Rev. D {\bf 61} 064018 (2000).
\href{https://doi.org/10.1103/PhysRevD.61.064018}{https://doi.org/10.1103/PhysRevD.61.064018}

\bibitem{AndGla1}Andersson, N., Glampedakis, K.:
Superradiance resonance cavity outside rapidly rotating black holes.
Phys. Rev. Lett. {\bf 84} 4537 (2000).
\href{https://doi.org/10.1103/PhysRevLett.84.4537}{https://doi.org/10.1103/PhysRevLett.84.4537}

\bibitem{AndGla2}Andersson, N., Glampedakis, K.:
Late-time dynamics of rapidly rotating black holes.
Phys. Rev. D {\bf 64} 104021 (2001).
\href{https://doi.org/10.1103/PhysRevD.64.104021}{https://doi.org/10.1103/PhysRevD.64.104021}

\bibitem{Poisson}Poisson, E.:
Radiative falloff of a scalar field in a weakly curved spacetime without symmetries.
Phys. Rev. D \textbf{66} 044008 (2002).
\href{https://doi.org/10.1103/PhysRevD.66.044008}{https://doi.org/10.1103/PhysRevD.66.044008}

\bibitem{Scheel}Scheel, M. A., \textit{et al.}:
3D simulations of linearized scalar fields in Kerr spacetime.
Phys. Rev. D \textbf{69} 104006 (2004).
\href{https://doi.org/10.1103/PhysRevD.69.104006}{https://doi.org/10.1103/PhysRevD.69.104006}

\bibitem{SchEriBurKidPfeTeu}Scheel, M. A., Erickcek, A. L., Burko, L. M., Kidder, L. E., Pfeiffer, H. P., Teukolsky, S. A.:
3D simulations of linearized scalar fields in Kerr spacetime.
Phys. Rev. D {\bf 69} 104006 (2004).
\href{https://doi.org/10.1103/PhysRevD.69.104006}{https://doi.org/10.1103/PhysRevD.69.104006}

\bibitem{PazLou}Pazos-Avalos, E., Lousto, C. O.:
Numerical integration of the Teukolsky equation in the time domain.
Phys. Rev. D {\bf 72} 084022 (2005).
\href{https://doi.org/10.1103/PhysRevD.72.084022}{https://doi.org/10.1103/PhysRevD.72.084022}

\bibitem{FinKamSmoYau}Finster, F., Kamran, N., Smoller, J., Yau, S.-T.:
Decay of solutions of the wave equation in the Kerr geometry.
Comm. Math. Phys. {\bf 264} 465-503 (2006).
\href{https://doi.org/10.1007/s00220-006-1525-8}{https://doi.org/10.1007/s00220-006-1525-8}
Erratum: Comm. Math. Phys. {\bf 280} 563-573 (2008).
\href{https://doi.org/10.1007/s00220-008-0458-9}{https://doi.org/10.1007/s00220-008-0458-9}


\bibitem{TigKidTeu}Tiglio, M., Kidder, L., Teukolsky, S.:
High accuracy simulations of Kerr tails: coordinate dependence and higher multipoles.
Class. Quantum Grav. \textbf{25} 105022 (2008).
\href{https://doi.org/10.1088/0264-9381/25/10/105022}{https://doi.org/10.1088/0264-9381/25/10/105022}

\bibitem{GlePriPul}Gleiser, R. J., Price, R. H., Pullin, J.:
Late-time tails in the Kerr spacetime.
Class. Quantum Grav. \textbf{25} 072001 (2008).
\href{https://doi.org/10.1088/0264-9381/25/7/072001}{https://doi.org/10.1088/0264-9381/25/7/072001}

\bibitem{BurKha3}Burko, L. M., Khanna, G.:
Late-time Kerr tails revisited.
Class. Quantum Grav. \textbf{26} 015014 (2009).
\href{https://doi.org/10.1088/0264-9381/26/1/015014}{https://doi.org/10.1088/0264-9381/26/1/015014}

\bibitem{BurKha}Burko, L. M., Khanna, G.:
Late-time Kerr tails: generic and non-generic initial data sets, 'up' modes, and superposition.
Class. Quantum Grav. \textbf{28} 025012 (2011).
\href{https://doi.org/10.1088/0264-9381/28/2/025012}{https://doi.org/10.1088/0264-9381/28/2/025012}

\bibitem{ZenTig}Zenginoglu, A., Tiglio, M.:
Spacelike matching to null infinity.
Phys. Rev. D \textbf{80} 024044 (2009).
\href{https://doi.org/10.1103/PhysRevD.80.024044}{https://doi.org/10.1103/PhysRevD.80.024044}

\bibitem{DafRod}Dafermos, M., Rodnianski, I.:
Decay for solutions of the wave equation on Kerr exterior spacetimes I-II: The cases $|a| << M$ or axisymmetry.
arXiv:1010.5132 [gr-qc].
\href{https://doi.org/10.48550/arXiv.1010.5132}{https://doi.org/10.48550/arXiv.1010.5132}

\bibitem{Luk}Luk, J.:
A vector field method approach to improved decay for solutions to the
wave equation on a slowly rotating Kerr black hole.
Anal. PDE {\bf 5} 553-623 (2012).
\href{http://dx.doi.org/10.2140/apde.2012.5.553}{http://dx.doi.org/10.2140/apde.2012.5.553}

\bibitem{Jasiulek}Jasiulek, M.:
Hyperboloidal slices for the wave equation of Kerr-Schild metrics and numerical applications.
Class. Quantum Grav. 29 015008 (2012).
\href{https://doi.org/10.1088/0264-9381/29/1/015008}{https://doi.org/10.1088/0264-9381/29/1/015008}

\bibitem{HBB}Harms, E., Bernuzzi, S., Br\"ugmann, B.:
Numerical solution of the 2+1 Teukolsky equation on a hyperboloidal and horizon penetrating foliation of Kerr and application to late-time decays.
Class. Quantum Grav. {\bf 30} 115013 (2013).
\href{https://doi.org/10.1088/0264-9381/30/11/115013}{https://doi.org/10.1088/0264-9381/30/11/115013}

\bibitem{Khanna}Khanna, G.:
High-precision numerical simulations on a CUDA GPU: Kerr black hole tails.
J. Sci. Comput. {\bf 56} 366-380 (2013).
\href{https://doi.org/10.1007/s10915-012-9679-3}{https://doi.org/10.1007/s10915-012-9679-3}

\bibitem{SpiKha}Spilhaus, T., Khanna, G.:
Brief note on high-multipole Kerr tails.
arXiv:1312.5210 [gr-qc].
\href{https://doi.org/10.48550/arXiv.1312.5210}{https://doi.org/10.48550/arXiv.1312.5210}

\bibitem{YanZimZen}Yang, H., Zimmerman, A., Zenginoglu, A., Zhang, F., Berti, E., Chen, Y.:
Quasinormal modes of nearly extremal Kerr spacetimes: Spectrum bifurcation and power-law ringdown.
Phys. Rev. D {\bf 88} 044047 (2013).
\href{https://doi.org/10.1103/PhysRevD.88.044047}{https://doi.org/10.1103/PhysRevD.88.044047}

\bibitem{Tataru2013}Tataru, D.:
Local decay of waves on asymptotically flat stationary space-times.
Am. J. Math. {\bf 135} 361-401 (2013).
\href{https://doi.org/10.1353/ajm.2013.0012}{https://doi.org/10.1353/ajm.2013.0012}

\bibitem{HolWal}Hollands, S., Wald, R. M.:
Stability of black holes and black branes.
Commun. Math. Phys. {\bf 321} 629-680 (2013).
\href{https://doi.org/10.1007/s00220-012-1638-1}{https://doi.org/10.1007/s00220-012-1638-1}

\bibitem{MacAns}Macedo, R. P., Ansorg, M.:
Axisymmetric fully spectral code for hyperbolic equations.
J. Computational Phys. \textbf{276} 357-379 (2014).
\href{https://doi.org/10.1016/j.jcp.2014.07.040}{https://doi.org/10.1016/j.jcp.2014.07.040}

\bibitem{ZenKhaBur}Zenginoglu, A., Khanna, G., Burko, L. M.:
Intermediate behavior of Kerr tails.
Gen. Relativ. Gravit. \textbf{46} 1672 (2014).
\href{https://doi.org/10.1007/s10714-014-1672-8}{https://doi.org/10.1007/s10714-014-1672-8}

\bibitem{BurKha4}Burko, L. M., Khanna, G.:
Mode coupling mechanism for late-time Kerr tails.
Phys. Rev. D \textbf{89} 044037 (2014).
\href{https://doi.org/10.1103/PhysRevD.89.044037}{https://doi.org/10.1103/PhysRevD.89.044037}

\bibitem{DaiAus1}Dain, S., Gentile de Austria, I.:
On the linear stability of the extreme Kerr black hole under axially symmetric perturbations.
Class. Quantum. Grav. {\bf 31} 195009 (2014).
\href{https://doi.org/10.1088/0264-9381/31/19/195009}{https://doi.org/10.1088/0264-9381/31/19/195009}

\bibitem{DaiAus2}Dain, S., Gentile de Austria, I.:
Bounds for axially symmetric linear perturbations for the extreme Kerr black hole.
Class. Quantum. Grav. {\bf 32} 135010 (2015).
\href{https://doi.org/10.1088/0264-9381/32/13/135010}{https://doi.org/10.1088/0264-9381/32/13/135010}

\bibitem{AndBlu}Andersson, L., Blue, P.:
Uniform energy bound and asymptotics for the Maxwell field on a slowly rotating Kerr black hole exterior.
J. Hyperbolic Differ. Equ. {\bf 12} 689-743 (2015).
\href{https://doi.org/10.1142/S0219891615500204}{https://doi.org/10.1142/S0219891615500204}

\bibitem{YSR}Shlapentokh-Rothman, Y.:
Quantitative mode stability for the wave equation on the Kerr spacetime.
Ann. Henri Poincare {\bf 16} 289-345 (2015).
\href{https://doi.org/10.1007/s00023-014-0315-7}{https://doi.org/10.1007/s00023-014-0315-7}

\bibitem{DafRodShl}Dafermos, M., Rodnianski, I., Shlapentokh-Rothman, Y.:
Decay for solutions of the wave equation on Kerr exterior spacetimes III: The full subextremal case $|a| < M$.
Annals of Mathematics {\bf 183} no. 3 787-913 (2016).
\href{https://doi.org/10.4007/annals.2016.183.3.2}{https://doi.org/10.4007/annals.2016.183.3.2}

\bibitem{CasGraZim}Casals, M., Gralla, S. E., Zimmerman, P.:
Horizon instability of extremal Kerr black holes: Nonaxisymmetric modes and enhanced growth rate.
Phys. Rev. D {\bf 94} 064003 (2016).
\href{https://doi.org/10.1103/PhysRevD.94.064003}{https://doi.org/10.1103/PhysRevD.94.064003}

\bibitem{CasKavOtt}Casals, M., Kavanagh, C., Ottewill, A. C.:
High-order late-time tail in a Kerr spacetime.
Phys. Rev. D {\bf 94} 124053 (2016).
\href{https://doi.org/10.1103/PhysRevD.94.124053}{https://doi.org/10.1103/PhysRevD.94.124053}

\bibitem{ThuKhaPri}Thuestad, I., Khanna, G., Price, R. H.:
Scalar fields in black hole spacetimes.
Phys. Rev. D \textbf{96} 024020 (2017).
\href{https://doi.org/10.1103/PhysRevD.96.024020}{https://doi.org/10.1103/PhysRevD.96.024020}

\bibitem{FinSmo}Finster, F., Smoller, J.:
Linear stability of the non-extreme Kerr black hole.
Adv. Theor. Math. Phys. {\bf 21} 1991-2085 (2017).
\href{https://doi.org/10.4310/ATMP.2017.v21.n8.a4}{https://doi.org/10.4310/ATMP.2017.v21.n8.a4}

\bibitem{MetTatToh}Metcalfe, J., Tataru, D., Tohaneanu, M.:
Pointwise decay for the Maxwell field on black hole spacetimes.
Adv. Math. {\bf 316} 53-93 (2017).
\href{https://doi.org/10.1016/j.aim.2017.05.024}{https://doi.org/10.1016/j.aim.2017.05.024}

\bibitem{GraZim}Gralla, S. E., Zimmerman, P.:
Critical exponents of extremal Kerr perturbations.
Class. Quantum Grav. {\bf 35} 095002 (2018).
\href{https://doi.org/10.1088/1361-6382/aab140}{https://doi.org/10.1088/1361-6382/aab140}

\bibitem{BurKha5}Burko L. M., Khanna, G.:
Linearized stability of extreme black holes.
Phys. Rev. D {\bf 97} 061502(R) (2018)
\href{https://doi.org/10.1103/PhysRevD.97.061502}{https://doi.org/10.1103/PhysRevD.97.061502}

\bibitem{PraWal}Prabhu, K., Wald, R. M.:
Stability of stationary-axisymmetric black holes in vacuum general relativity to axisymmetric electromagnetic perturbations.
Class. Quantum Grav. {\bf 35} 015009 (2018).
\href{https://doi.org/10.1088/1361-6382/aa95ef}{https://doi.org/10.1088/1361-6382/aa95ef}

\bibitem{MaPHD}Ma, S.:
Analysis of Teukolsky equations on slowly rotating Kerr spacetimes.
Ph.D. thesis, Universit\"at Potsdam, 2018

\bibitem{DafHolRod}Dafermos, M., Holzegel, G., Rodnianski, I.:
Boundedness and decay for the Teukolsky equation on Kerr spacetimes I: The case $|a| << M$.
Ann. PDE {\bf 5} 2 (2019).
\href{https://doi.org/10.1007/s40818-018-0058-8}{https://doi.org/10.1007/s40818-018-0058-8}

\bibitem{AndBacBluMa}Andersson, L., B\"ackdahl, T., Blue, P., Ma, S.:
Stability for linearized gravity on the Kerr spacetime.
arXiv:1903.03859 [math.AP].
\href{https://doi.org/10.48550/arXiv.1903.03859}{https://doi.org/10.48550/arXiv.1903.03859}


\bibitem{CasZim}Casals, M., Zimmerman, P.:
Perturbations of an extremal Kerr spacetime: Analytic framework and late-time tails.
Phys. Rev. D \textbf{100} 124027 (2019).
\href{https://doi.org/10.1103/PhysRevD.100.124027}{https://doi.org/10.1103/PhysRevD.100.124027}

\bibitem{Gudapati}Gudapati, N.:
A conserved energy for axially symmetric Newman-Penrose-Maxwell scalars on Kerr black holes.
Proc. Roy. Soc. A {\bf 475} (2019) no. 2221, 20180686.
\href{https://doi.org/10.1098/rspa.2018.0686}{https://doi.org/10.1098/rspa.2018.0686}

\bibitem{Ma}Ma, S.:
Almost Price's law in Schwarzschild and decay estimates in Kerr for Maxwell field.
J. Diff. Eq. {\bf 339} 1-89 (2022).
\href{https://doi.org/10.1016/j.jde.2022.08.021}{https://doi.org/10.1016/j.jde.2022.08.021}

\bibitem{Ma2}Ma, S.:
Uniform energy bound and Morawetz estimate for extreme components of spin fields in the exterior of a slowly rotating Kerr black hole I:
Maxwell field.
Ann. Henri Poincare {\bf 21} 815-863 (2020).
\href{https://doi.org/10.1007/s00023-020-00884-7}{https://doi.org/10.1007/s00023-020-00884-7}

\bibitem{Ma3}Ma, S.:
Uniform energy bound and Morawetz estimate for extreme components of spin fields in the exterior of a slowly rotating Kerr black hole II:
linearized gravity.
Commun. Math. Phys. {\bf 377} 2489-2551 (2020).
\href{https://doi.org/10.1007/s00220-020-03777-2}{https://doi.org/10.1007/s00220-020-03777-2}

\bibitem{Costa}Teixeira da Costa, R.:
Mode stability for the Teukolsky equation on extremal and subextremal Kerr spacetimes.
Commun. Math. Phys. {\bf 378} 705-781 (2020).
\href{https://doi.org/10.1007/s00220-020-03796-z}{https://doi.org/10.1007/s00220-020-03796-z}

\bibitem{ShlCos}Shlapentokh-Rothman, Y., Teixeira da Costa, R.:
Boundedness and decay for the Teukolsky equation on Kerr in the full subextremal range $|a|<M$: frequency space analysis.
arXiv:2007.07211 [gr-qc].
\href{https://doi.org/10.48550/arXiv.2007.07211}{https://doi.org/10.48550/arXiv.2007.07211}

\bibitem{GudMon}Moncrief, V., Gudapati, N.:
A positive-definite energy functional for the axisymmetric perturbations of Kerr-Newman black holes.
arXiv:2105.12632 [gr-qc].
\href{https://doi.org/10.48550/arXiv.2105.12632}{https://doi.org/10.48550/arXiv.2105.12632}

\bibitem{RipLouGioPre}Ripley, J. L., Loutrel, N., Giorgi, E., Pretorius, F.:
Numerical computation of second-order vacuum perturbations of Kerr black holes.
Phys. Rev. D {\bf 103} 104018 (2021).
\href{https://doi.org/10.1103/PhysRevD.103.104018}{https://doi.org/10.1103/PhysRevD.103.104018}

\bibitem{AngAreGaj}Angelopoulos, Y., Aretakis, S., Gajic, D.:
Late time tails and mode coupling of linear waves on Kerr spacetimes.
Adv. Math. {\bf 417} 108939 (2023).
\href{https://doi.org/10.1016/j.aim.2023.108939}{https://doi.org/10.1016/j.aim.2023.108939}

\bibitem{MaZhang}Ma, S., Zhang, L.:
Sharp decay for Teukolsky equation in Kerr spacetimes.
Commun. Math. Phys. {\bf 401} 333-434 (2023).
\href{https://doi.org/10.1007/s00220-023-04640-w}{https://doi.org/10.1007/s00220-023-04640-w}

\bibitem{HafHinVas}H\"afner, D., Hintz, P., Vasy, A.:
Linear stability of slowly rotating Kerr black holes.
Invent. math. 223, 1227-1406 (2021).
\href{https://doi.org/10.1007/s00222-020-01002-4}{https://doi.org/10.1007/s00222-020-01002-4}

\bibitem{GajKeh}Gajic, D., Kehrberger, L. M. A.:
On the relation between asymptotic charges, the failure of peeling and late-time tails.
Class. Quantum Grav. {\bf 39} 195006 (2022).
\href{https://doi.org/10.1088/1361-6382/ac8863}{https://doi.org/10.1088/1361-6382/ac8863}

\bibitem{Hintz}Hintz, P.:
A sharp version of Price's law for wave decay on asymptotically flat spacetimes.
Commun. Math. Phys. 389, 491-542 (2022).
\href{https://doi.org/10.1007/s00220-021-04276-8}{https://doi.org/10.1007/s00220-021-04276-8}

\bibitem{FieGotGra}Field, S.E., Gottlieb, S., Grant, Z.J. et al.:
A GPU-accelerated mixed-precision WENO method for extremal black hole and gravitational wave physics computations.
Commun. Appl. Math. Comput. {\bf 5} 97-115 (2023).
\href{https://doi.org/10.1007/s42967-021-00129-2}{https://doi.org/10.1007/s42967-021-00129-2}

\bibitem{Millet}Millet, P.:
Optimal decay for solutions of the Teukolsky equation on the Kerr metric for the full subextremal range $|a|<M$.
arXiv:2302.06946 [math.AP].
\href{https://doi.org/10.48550/arXiv.2302.06946}{https://doi.org/10.48550/arXiv.2302.06946}


\bibitem{RegWhe}Regge, T., Wheeler, J. A.:
Stability of a Schwarzschild singularity.
Phys. Rev. {\bf 108}  1063 (1957).
\href{https://doi.org/10.1103/PhysRev.108.1063}{https://doi.org/10.1103/PhysRev.108.1063}

\bibitem{Vishveshwara}Vishveshwara, C. V.:
Stability of the Schwarzschild metric.
Phys. Rev. D {\bf 1}  2870-2879 (1970).
\href{https://doi.org/10.1103/PhysRevD.1.2870}{https://doi.org/10.1103/PhysRevD.1.2870}

\bibitem{Moncrief1}Moncrief, V.:
Gauge invariant perturbations of Reissner-Nordstrom black holes.
Phys. Rev. D {\bf 12} 1526-1537 (1974).
\href{https://doi.org/10.1103/PhysRevD.12.1526}{https://doi.org/10.1103/PhysRevD.12.1526}

\bibitem{Moncrief2}Moncrief, V.:
Odd-parity stability of a Reissner-Nordstrom black hole.
Phys. Rev. D {\bf 9} 2707-2709 (1974).
\href{https://doi.org/10.1103/PhysRevD.9.2707}{https://doi.org/10.1103/PhysRevD.9.2707}

\bibitem{Moncrief3}Moncrief, V.:
Stability of Reissner-Nordstrom black holes.
Phys. Rev. D {\bf 10} 1057-1059 (1974).
\href{https://doi.org/10.1103/PhysRevD.10.1057}{https://doi.org/10.1103/PhysRevD.10.1057}

\bibitem{Wald}Wald, R. M.:
Note on the stability of the Schwarzschild metric.
J. Math. Phys. {\bf 20}  1056-1058 (1979).
\href{https://doi.org/10.1063/1.524181}{https://doi.org/10.1063/1.524181}

\bibitem{KayWald}Kay, B. S., Wald, R. M.:
Linear stability of Schwarzschild under perturbations which are
non-vanishing on the bifurcation 2-sphere.
Class. Quantum Grav. {\bf 4}  893 (1987).
\href{https://doi.org/10.1088/0264-9381/4/4/022}{https://doi.org/10.1088/0264-9381/4/4/022}

\bibitem{Leaver}Leaver, E. W.:
Spectral decomposition of the perturbation response of the Schwarzschild geometry.
Phys. Rev. D \textbf{34} 384 (1986).
\href{https://doi.org/10.1103/PhysRevD.34.384}{https://doi.org/10.1103/PhysRevD.34.384};
Phys. Rev. D \textbf{38} 725 (1988) (Erratum).
\href{https://doi.org/10.1103/PhysRevD.38.725}{https://doi.org/10.1103/PhysRevD.38.725}

\bibitem{GunPriPul}Gundlach, C., Price, R. H., Pullin, J.:
Late-time behavior of stellar collapse and explosions. I. Linearized perturbations.
Phys. Rev. D {\bf 49}  883 (1994).
\href{https://doi.org/10.1103/PhysRevD.49.883}{https://doi.org/10.1103/PhysRevD.49.883}

\bibitem{Barack2}Barack, L.:
Late time dynamics of scalar perturbations outside black holes. II. Schwarzschild geometry.
Phys. Rev. D {\bf 59}  044017 (1999).
\href{https://doi.org/10.1103/PhysRevD.59.044017}{https://doi.org/10.1103/PhysRevD.59.044017}

\bibitem{KarSwiMal}Karkowski, J., Swierczynski, Z., Malec, E.:
Comments on tails in Schwarzschild spacetimes.
Class. Quantum Grav. \textbf{21} 1303 (2004).
\href{https://doi.org/10.1088/0264-9381/21/6/002}{https://doi.org/10.1088/0264-9381/21/6/002}

\bibitem{PriBur}Price, R. H., Burko, L. M.:
Late time tails from momentarily stationary, compact initial data in Schwarzschild spacetimes.
Phys. Rev. D \textbf{70} 084039 (2004).
\href{https://doi.org/10.1103/PhysRevD.70.084039}{https://doi.org/10.1103/PhysRevD.70.084039}

\bibitem{CalGunHil}Calabrese, G., Gundlach, C., Hilditch, D.:
Asymptotically null slices in numerical relativity: mathematical analysis and spherical wave equation tests.
Class. Quantum Grav. \textbf{23} 4829 (2006).
\href{https://doi.org/10.1088/0264-9381/23/15/004}{https://doi.org/10.1088/0264-9381/23/15/004}

\bibitem{BerNagPie}Bernuzzi, S., Nagar, A., De Pietri, R.:
Dynamical excitation of space-time modes of compact objects.
Phys. Rev. D {\bf 77} 044042 (2008).
\href{https://doi.org/10.1103/PhysRevD.77.044042}{https://doi.org/10.1103/PhysRevD.77.044042}

\bibitem{Blue}Blue, P.:
Decay of the Maxwell field on the Schwarzschild manifold.
J. Hyperbolic Differ. Equ. 5(4) 807-856 (2008).
\href{https://doi.org/10.1142/S0219891608001714}{https://doi.org/10.1142/S0219891608001714}

\bibitem{ZenNunHus}Zenginoglu, A., Nunez, D., Husa, S.:
Gravitational perturbations of Schwarzschild spacetime at null infinity and the hyperboloidal initial value problem.
Class. Quantum Grav. {\bf 26} 035009 (2009).
\href{https://doi.org/10.1088/0264-9381/26/3/035009}{https://doi.org/10.1088/0264-9381/26/3/035009}

\bibitem{ZenKid}Zenginoglu, A., Kidder, L. E.:
Hyperboloidal evolution of test fields in three spatial dimensions.
Phys. Rev. D {\bf 81}  124010 (2010).
\href{https://doi.org/10.1103/PhysRevD.81.124010}{https://doi.org/10.1103/PhysRevD.81.124010}

\bibitem{DonSchSof1}Donninger, R., Schlag, W., Soffer, A.:
A proof of Price's Law on Schwarzschild black hole manifolds for all angular momenta.
Adv. Math. {\bf 226} 484-540 (2011).
\href{https://doi.org/10.1016/j.aim.2010.06.026}{https://doi.org/10.1016/j.aim.2010.06.026}

\bibitem{DonSchSof2}Donninger, R., Schlag, W., Soffer, A.:
On pointwise decay of linear waves on a Schwarzschild black hole background.
Comm. Math. Phys. {\bf 309} 51 (2012).
\href{https://doi.org/10.1007/s00220-011-1393-8}{https://doi.org/10.1007/s00220-011-1393-8}

\bibitem{Dotti1}Dotti, G.:
Nonmodal linear stability of the Schwarzschild black hole.
Phys. Rev. Lett. {\bf 112} 191101 (2014).
\href{https://doi.org/10.1103/PhysRevLett.112.191101}{https://doi.org/10.1103/PhysRevLett.112.191101}

\bibitem{Ghanem}Ghanem, S.:
On uniform decay of the Maxwell fields on black hole space-times.
arXiv:1409.8040 [math.AP].
\href{https://doi.org/10.48550/arXiv.1409.8040}{https://doi.org/10.48550/arXiv.1409.8040}

\bibitem{SteTat}Sterbenz, J., Tataru, D.:
Local energy decay for Maxwell fields part I: Spherically symmetric black-hole backgrounds.
International Mathematics Research Notices 2015, Vol. 2015 Issue 11, 3298-3342 

\bibitem{CasOtt}Casals, M., Ottewill, A. C.:
High-order tail in Schwarzschild spacetime.
Phys. Rev. D {\bf 92} 124055 (2015).
\href{https://doi.org/10.1103/PhysRevD.92.124055}{https://doi.org/10.1103/PhysRevD.92.124055}

\bibitem{AndBacBlu}Andersson, L., B\"ackdahl, T., Blue, P.:
Decay of solutions to the Maxwell equation on the Schwarzschild background.
Class. Quantum Grav. {\bf 33} 085010 (2016).
\href{https://doi.org/10.1088/0264-9381/33/8/085010}{https://doi.org/10.1088/0264-9381/33/8/085010}

\bibitem{HunKelWan}Hung, Pei-Ken, Keller, J., Wang, Mu-Tao:
Linear stability of Schwarzschild spacetime: decay of metric coefficients.
J. Differential Geom. {\bf 116} 481-541 (2020).
\href{https://doi.org/10.4310/jdg/1606964416}{https://doi.org/10.4310/jdg/1606964416}

\bibitem{Pasqualotto}Pasqualotto, F.:
The spin $\pm 1$ Teukolsky equations and the Maxwell system on Schwarzschild.
Ann. Henri Poincare {\bf 20} 1263-1323 (2019).
\href{https://doi.org/10.1007/s00023-019-00785-4}{https://doi.org/10.1007/s00023-019-00785-4}

\bibitem{DafHolRod2}Dafermos, M., Holzegel, G., Rodnianski, I.:
The linear stability of the Schwarzschild solution to gravitational perturbations.
Acta Math. {\bf 222}  1-214 (2019).
\href{https://dx.doi.org/10.4310/ACTA.2019.v222.n1.a1}{https://dx.doi.org/10.4310/ACTA.2019.v222.n1.a1}

\bibitem{AndBluWan}Andersson, L., Blue, P., Wang, J.:
Morawetz estimate for linearized gravity in Schwarzschild.
Ann. Henri Poincare {\bf 21} 761-813 (2020).
\href{https://doi.org/10.1007/s00023-020-00886-5}{https://doi.org/10.1007/s00023-020-00886-5}


\bibitem{Giorgi1}Giorgi, E.:
The linear stability of Reissner-Nordstrom spacetime for small charge.
Ann. PDE {\bf 6}  8 (2020).
\href{https://doi.org/10.1007/s40818-020-00082-y}{https://doi.org/10.1007/s40818-020-00082-y}

\bibitem{Giorgi2}Giorgi, E.:
The linear stability of Reissner-Nordstrom spacetime: the full subextremal range $|Q|<M$.
Commun. Math. Phys. {\bf 380}  1313-1360 (2020).
\href{https://doi.org/10.1007/s00220-020-03893-z}{https://doi.org/10.1007/s00220-020-03893-z}

\bibitem{MaZhang2}Ma, S., Zhang, L.:
Price's law for spin fields on a Schwarzschild background.
Ann. PDE {\bf 8} 25 (2022).
\href{https://doi.org/10.1007/s40818-022-00139-0}{https://doi.org/10.1007/s40818-022-00139-0}

\bibitem{AngAreGaj2}Angelopoulos, Y., Aretakis, S., Gajic, D.:
Price's law and precise late-time asymptotics for subextremal Reissner-Nordstrom black holes.
Ann. Henri Poincare {\bf 24} 3215-3287 (2023).
\href{https://doi.org/10.1007/s00023-023-01328-8}{https://doi.org/10.1007/s00023-023-01328-8}




\bibitem{DafRod-artb}Dafermos, M., Rodnianski, I.:
A note on boundary value problems for black hole evolutions.
arXiv:gr-qc/0403034.
\href{https://doi.org/10.48550/arXiv.gr-qc/0403034}{https://doi.org/10.48550/arXiv.gr-qc/0403034}




\bibitem{Penrose1}Penrose, R.:
Asymptotic properties of fields and space-times.
Phys. Rev. Lett. \textbf{10} 66 (1963).
\href{https://doi.org/10.1103/PhysRevLett.10.66}{https://doi.org/10.1103/PhysRevLett.10.66}

\bibitem{Penrose3}Penrose, R.,
in Relativity, Groups and Topology, edited by DeWitt, C. and DeWitt, B. (Gordon and Breach, New York, 1964), p. 565

\bibitem{Moncrief}Moncrief, V.:
Conformally regular ADM evolution equations, 2000,
Proceedings of the Workshop on
Mathematical Issues in Numerical Relativity, Santa Barbara

\bibitem{FR1}Fodor, G., R\'acz, I.:
What does a strongly excited 't Hooft-Polyakov magnetic monopole do?
Phys. Rev. Lett. \textbf{92} 151801 (2004).
\href{https://doi.org/10.1103/PhysRevLett.92.151801}{https://doi.org/10.1103/PhysRevLett.92.151801}

\bibitem{FR2}Fodor G., R\'acz, I.:
Numerical investigation of highly excited magnetic monopoles in SU(2) Yang-Mills-Higgs theory.
Phys. Rev. D \textbf{77} 025019 (2008).
\href{https://doi.org/10.1103/PhysRevD.77.025019}{https://doi.org/10.1103/PhysRevD.77.025019}

\bibitem{Zenginoglu}Zenginoglu, A.:
Hyperboloidal foliations and scri-fixing.
Class. Quantum Grav. \textbf{25} 145002 (2008).
\href{https://doi.org/10.1088/0264-9381/25/14/145002}{https://doi.org/10.1088/0264-9381/25/14/145002}

\bibitem{MacJarAns}Macedo, R. P., Jaramillo, J. L., Ansorg, M.:
Hyperboloidal slicing approach to quasinormal mode expansions: The Reissner-Nordstr\"om case.
Phys. Rev. D \textbf{98} 124005 (2018).
\href{https://doi.org/10.1103/PhysRevD.98.124005}{https://doi.org/10.1103/PhysRevD.98.124005}

\bibitem{Macedo}Macedo, R. P.:
Hyperboloidal framework for the Kerr spacetime.
Class. Quantum Grav. \textbf{37} 065019 (2020).
\href{https://doi.org/10.1088/1361-6382/ab6e3e}{https://doi.org/10.1088/1361-6382/ab6e3e}




\bibitem{CLR}Csizmadia, P., L\'aszl\'o, A., R\'acz, I.:
On the use of multipole expansion in time evolution of
non-linear dynamical systems and some surprises related to superradiance.
Class. Quantum Grav. {\bf 30} 015010 (2013).
\href{https://doi.org/10.1088/0264-9381/30/1/015010}{https://doi.org/10.1088/0264-9381/30/1/015010}


\bibitem{CroFac}Crossman, R. G., Fackerell, E. D.:
Electrovac perturbations of rotating black holes.
in: Edwards, C. (eds) Gravitational Radiation, Collapsed Objects and Exact Solutions,
Lecture Notes in Physics {\bf 124} (1980). Springer, Berlin, Heidelberg.
\href{https://doi.org/10.1007/3-540-09992-1_118}{https://doi.org/10.1007/3-540-09992-1\_118}


\bibitem{BinCheJanRuf}Bini, D., Cherubini, C., Jantzen, R. T., Ruffini, R.:
Teukolsky master equation: de Rham wave equation for gravitational and electromagnetic fields in vacuum.
Prog. Theor. Phys. {\bf 107} 967-992 (2002).
\href{https://doi.org/10.1143/PTP.107.967}{https://doi.org/10.1143/PTP.107.967}

\bibitem{AndAks}Andersson, L., Aksteiner, S.:
Linearized gravity and gauge conditions.
Class. Quantum Grav. {\bf 28} 065001 (2011).
\href{https://doi.org/10.1088/0264-9381/28/6/065001}{https://doi.org/10.1088/0264-9381/28/6/065001}

\bibitem{Aksteiner}Aksteiner, S.: \emph{Geometry and Analysis on Black Hole Spacetimes},
PhD dissertation, Gottfried Wilhelm Leibniz Universit\"at Hannover (2014)
\href{https://d-nb.info/1057896721/34}{d-nb.info/1057896721/34}

\bibitem{JezSmo}Jezierski, J., Smolka, T.:
A geometric description of Maxwell field in a Kerr spacetime.
Class. Quantum Grav. {\bf 33} 125035 (2016).
\href{https://doi.org/10.1088/0264-9381/33/12/125035}{https://doi.org/10.1088/0264-9381/33/12/125035}

\bibitem{Araneda}Araneda, B.:
Symmetry operators and decoupled equations for linear fields on black hole spacetimes.
Class. Quantum Grav. {\bf 34} 035002 (2017).
\href{https://doi.org/10.1088/1361-6382/aa51ff}{https://doi.org/10.1088/1361-6382/aa51ff}


\bibitem{RT}R\'acz, I., T\'oth, G. Z.:
Numerical investigation of the late-time Kerr tails.
Class. Quantum Grav. {\bf 28} 195003 (2011)
\href{https://doi.org/10.1088/0264-9381/28/19/195003}{https://doi.org/10.1088/0264-9381/28/19/195003}

\bibitem{Toth}T\'oth, G. Z.:
Noether currents for the Teukolsky master equation.
Class. Quantum Grav. {\bf 35} 185009 (2018).
\href{https://doi.org/10.1088/1361-6382/aad712}{https://doi.org/10.1088/1361-6382/aad712}

\bibitem{CRT}Csuk\'as, K., R\'acz, I., T\'oth, G. Z.:
Numerical investigation of the dynamics of linear spin s fields
on a Kerr background: Late-time tails of spin s = $\pm 1$,  $\pm 2$ fields.
Phys. Rev. D {\bf 100} 104025 (2019).
\href{https://doi.org/10.1103/PhysRevD.100.104025}{https://doi.org/10.1103/PhysRevD.100.104025}


\bibitem{ABl}Andersson, L., Blue, P.:
Hidden symmetries and decay for the wave equation on the Kerr spacetime.
Ann. of Math. {\bf 182}(3) 787-853 (2015).
\href{https://doi.org/10.4007/annals.2015.182.3.1}{https://doi.org/10.4007/annals.2015.182.3.1}

\bibitem{AndBacBlu-cl}Andersson, L., B\"ackdahl, T., Blue, P.:
Spin geometry and conservation laws in the Kerr spacetime.
Surveys in Differential Geometry {\bf 20} 183-226 (2015).
\href{https://dx.doi.org/10.4310/SDG.2015.v20.n1.a8}{https://dx.doi.org/10.4310/SDG.2015.v20.n1.a8}
(in ``One hundred years of general relativity'' edited by Lydia Bieri and Shing-Tung Yau)


\bibitem{ABB1}Andersson, L., B\"ackdahl, T., Blue, P.:
A new tensorial conservation law for Maxwell fields on the Kerr background.
J. Diff. Geom. {\bf 105} no.2 163 (2017).
\href{https://doi.org/10.4310/jdg/1486522812}{https://doi.org/10.4310/jdg/1486522812}


\bibitem{CR}Csuk\'as, K., R\'acz, I.:
Numerical investigation of the dynamics of linear spin s fields on a Kerr background. II. Superradiant scattering.
Phys. Rev. D {\bf 103} 084035 (2021).
\href{https://doi.org/10.1103/PhysRevD.103.084035}{https://doi.org/10.1103/PhysRevD.103.084035}


\bibitem{GrFl1}Grant, A. M., Flanagan, \'E. \'E.:
Conserved currents for electromagnetic fields in the Kerr spacetime.
Class. Quantum Grav. {\bf 37} 185021 (2020).
\href{https://doi.org/10.1088/1361-6382/ab995a}{https://doi.org/10.1088/1361-6382/ab995a}

\bibitem{GrFl2}Grant, A. M., Flanagan, \'E. \'E.:
A class of conserved currents for linearized gravity in the Kerr spacetime.
Class. Quantum Grav. {\bf 38} 055004 (2021).
\href{https://doi.org/10.1088/1361-6382/abc3f7}{https://doi.org/10.1088/1361-6382/abc3f7}

\bibitem{GreHolSbeTooZim}Green, S. R., Hollands, S., Sberna, L., Toomani, V., Zimmerman, P.:
Conserved currents for a Kerr black hole and orthogonality of quasinormal modes.
Phys. Rev. D {\bf 107} 064030 (2023).
\href{https://doi.org/10.1103/PhysRevD.107.064030}{https://doi.org/10.1103/PhysRevD.107.064030}


\bibitem{Gustetal}Gustafsson, B., Kreiss, H-O., Oliger, J.:
Time Dependent Problems and Difference Methods.
Pure and Applied Mathematics,
John Wiley and Sons, Inc., New York (1995)



\end{thebibliography}
\end{document}